\documentclass[referee]{raa}            % referee version: for submission
\usepackage{textcomp} % Provides the \micro symbol
\usepackage{graphicx,times}             %for PS/EPS graphics inclusion, new
\usepackage{natbib}
\usepackage{amssymb,amsmath}
\bibpunct{(}{)}{;}{a}{}{,}

\usepackage[a4paper]{geometry}
\usepackage[pagebackref=true]{hyperref}
\usepackage{tabularx}
\usepackage{booktabs}
\usepackage{longtable}
\usepackage{bm}
\usepackage{gensymb}
\usepackage[figuresright]{rotating}

\begin{document}

  \title{A Cartesian catalog of 30 million {\it Gaia} sources based on second-order and Monte Carlo error propagation}
  %\subtitle{A Cartesian catalog of {\it Gaia} sources}

   \volnopage{Vol.0 (20xx) No.0, 000--000}      %%preserved for Editor. DOn't remove!
   \setcounter{page}{1}          %%starting page, preserved for Editor. DOn't remove!

   \author{Luyao Zhang %(周爱英) %% Put your Chinese name in "( )" if you like. Note to open line 11 "\usepackage[UTF8]{ctex}"
      %\inst{1}
   \and Fabo Feng
      %\inst{1}
   \and Yicheng Rui
      %\inst{1}
    \and Guang-Yao Xiao
      %\inst{1}
   \and Wenting Wang
      %\inst{1}
   }


%%   \institute{School of Physics and Astronomy, Shanghai Jiao Tong University, 800 Dongchuan Road, Shanghai 200240, China\\
   %{\it luyoo.z@sjtu.edu.cn}\\
%%        \and
   \institute{State Key Laboratory of Dark Matter Physics, Tsung-Dao Lee Institute \& School of Physics and Astronomy, Shanghai Jiao Tong University, Shanghai 201210, China  \\
\vs\no
   {\small Received 20xx month day; accepted 20xx month day}}

\abstract{ Accurate measurements of stellar positions and velocities are crucial for studying galactic and stellar dynamics. We aim to create a Cartesian catalog from {\it Gaia} DR3 to serve as a high-precision database for further research using stellar coordinates and velocities. 
To avoid the negative parallax values, we select 31,129,169 sources in {\it Gaia} DR3 with radial velocity, where the fractional parallax error is less than 20\% ($0 < \sigma_\varpi/\varpi < 0.2$). 
To select the most accurate and efficient method of propagating mean and covariance, we use the Monte Carlo results with $10^7$ samples (MC7) as the benchmark, and compare the precision of linear, second-order, and Monte Carlo error propagation methods. By assessing the accuracy of propagated mean and covariance, we observe that second-order error propagation exhibits mean deviations of at most 0.5\% compared to MC7, with variance deviations of up to 10\%. Overall, this outperforms linear transformation. Though Monte Carlo method with $10^4$ samples (MC4) is an order of magnitude slower than second-order error propagation, its covariances propagation accuracy reaches 1\% when $\sigma_\varpi/\varpi$ is below 15\%. Consequently, we employ second-order error propagation to convert the mean astrometry and radial velocity into Cartesian coordinates and velocities in both equatorial and galactic systems for 30 million {\it Gaia} sources, and apply MC4 for covariance propagation. 
The Cartesian catalog and source code are provided for future applications in high-precision stellar and galactic dynamics. 
 }

   %\authorrunning{lL.Y. Zhang, E. Rodriguez \& B. J. Smith }            %author_head in even pages
   \titlerunning{A Cartesian catalog of {\it Gaia} sources }  % title_head in odd pages

   \maketitle
%% The author head (on even pages) and the title head (on odd pages) will be
%% automatically extracted from \author{} and \title{}. Whenever the title is too long,
%% you will be asked to supply a shorter one by inserting either \authorrunning{} or
%% \titlerunning{} before \maketitle. Anyway, you can specify your own heads.
%%
%%
%% Note: In the following text body of your manuscript, please note several differences from
%%       other major journals:
%% (1) \subsection{Please Capitalize the First Letter of Each Notional Word in Subsection Title}
%% (2) Please Capitalize the First Letter of Each Notional Word in all tables' captions
%
%________________________________________________ sections below
%
\section{Introduction}           %% first-level sections will be auto-capitalized
\label{sect:intro}
The demand for high-precision reconstruction of stellar orbits has gained prominence, particularly with the emergence of advanced astrometric missions like {\it Gaia} \citep{gaia16,gaia18}. {\it Gaia} Data Release 3 (DR3) offers the largest collection of all-sky spectrophotometry, radial velocities, and astrophysical parameters for stars. It is highly accurate in astrometric measurements for $\rm{G < 15~mag}$, with median position uncertainties ranging from 0.01 to 0.02 mas, median parallax uncertainties between 0.02 and 0.03 mas, median proper motion uncertainties from 0.02 to 0.03 mas/yr, and the radial velocity uncertainty of approximately 1 km/s \citep{gaia21, gaia23}. 
Such extraordinary precision in data has become a cornerstone in Galactic and stellar dynamics research, unlocking new possibilities for understanding the intricacies of celestial motion and providing unprecedented insights into the structures and dynamics of our galaxy.
%%change: $\sim$20,$\mu$as.

We can resolve individual stars in and around our host Galaxy, the Milky Way (MW). These include the stars in tidal debris or stellar streams, which are stripped by tidal forces from dwarf galaxies and globular clusters orbiting around the MW. In the field of galactic dynamics, the modeling of stellar streams often combines an approximate treatment of the stream formation to a high resolution simulation of galaxy formation. For streams whose progenitor galaxy or globular cluster are not fully disrupted by tides, the currently observed position and velocity of the progenitor can be integrated back in time with a fiducial MW potential model. It is then integrated forward in time, with tracer particles released from the two Lagrangian points to simulate the formation and evolution of the stream in the galactic tidal field. By comparing the simulated stream to the observation, one can constrain the underlying dark matter distribution \citep[e.g.][]{2012MNRAS.420.2700K,2014ApJ...795...94B,2015ApJ...803...80K,2017MNRAS.464..794G,2023MNRAS.524.2124P}. The modeling of stream orbit strongly depends on the initial condition set up by the Cartesian coordinates of the progenitors, and is often achieved in action and action-frequency space \citep[e.g.][]{2013MNRAS.433.1826S,2014ApJ...795...95B,2014MNRAS.443..423S,2016ApJ...833...31B}, hence sensitive to error propagation.

In the pursuit of understanding wide binaries, achieving precise error propagation is paramount, given the subtle velocity differences, often as minimal as $\sim$1\,kms$^{-1}$, exhibited by these binary systems. Harnessing the high-precision astrometric data generously provided by {\it Gaia} \citep{2016A&A...595A...1G}, millions of wide binaries are identified through the meticulous examination of their common proper motion and parallax \citep[e.g.,][]{El-Badry18, tian20, El-Badry21}. Particularly noteworthy is the utilization of the relative velocity and relative position angle of wide binaries to explore their eccentricity distribution \citep{hwang22}. Given that these investigations heavily rely on the accurate Cartesian coordinates and velocities of stars, the precise transformation of coordinates from {\it Gaia} astrometry, coupled with meticulous error propagation, becomes imperative. This is not only crucial for avoiding false positives in wide binary selection but also for enhancing the reliability of statistical studies concerning their properties.

Unlike wide binaries, stellar encounters encompass the serendipitous alignment of stars, occurring when they come into close proximity over a relatively short period. Notably, stellar encounters within the solar system significantly contribute to shaping the structure of the Oort Cloud \cite[e.g.,][]{sanchez01, 2004ASPC..323..371D, rickman08, RICKMAN2012124, 10.1093/mnras/stv2222, 2022A&A...664A.123D} . 
The exploration of slow and close encounters, involving interstellar objects like `Oumuamua, yields valuable insights into identifying their home systems and understanding their dynamical history \citep[e.g.,][]{Feng_2018_Oumuamua, Bailer_Jones_2018, Portegies_Zwart_2018, Loeb_2022}. Accurate stellar orbital integration proves essential in pinpointing stellar encounters, necessitating the precise propagation of stellar motion from an initial epoch to millions of years in the past or future. Given the nonlinear nature of stellar orbits over a million-year time scale, the encounter time and distance are highly sensitive to the initial Cartesian state of the stars \citep{dybczynski15}.

Recognizing the critical role of error propagation in numerous astrophysical applications, we undertake a comparative analysis of different methodologies for error propagation in the transformation of {\it Gaia} astrometry into Cartesian positions and velocities of stars. Specifically, we investigate the linear, second-order, and Monte Carlo (MC) error propagation methods. While advanced techniques such as the Kalman filter and unscented transformation \citep[e.g.,][]{smith62, schmidt66, julier04, Chen2017, michelotti2024comparisonuncertaintypropagationtechniques} are commonly applied in nonlinear-system tracking and navigation, second-order error propagation consistently achieves precision comparable to these approaches. This is exemplified by \cite{Feng_2018}, who conducted a comprehensive comparison of various methods in the orbital integration of stars. Our study reaffirms that second-order error propagation surpasses linear propagation in accuracy. Consequently, we use second-order error propagation to derive Cartesian positions and velocities in both Equatorial and Galactic coordinate systems from {\it Gaia} astrometry, and apply MC with $10^4$ samples for covariance conversion. This approach not only enhances the efficiency and accuracy of coordinate transformation but also establishes a robust foundation for research across diverse astrophysical fields.
%% change: Celestial-->Equatorial

The structure of this paper unfolds as follows: In section~\ref{data}, we provide a detailed overview of the research analysis, elucidating the data employed for error propagation. Section~\ref{method} delves into the techniques utilized for coordinate transformation and covariance propagation, encompassing MC, linear, and second-order error propagation methods. Transitioning to section~\ref{results}, we present and discuss the extensive outcomes derived from our methodology, focusing on approximately 30 million sources in the {\it Gaia} DR3. Lastly, section~\ref{conclusion} encapsulates a succinct discussion and conclusion.

%% Authors can give a citation as 'Michel et al. 1992'.
%% You may also use \cite, \citep and \citet for citation, and use Table~1 or Figure~1
%% and so forth. Using \ref and \label for cross-references of Tables/Figures
%% is a good way in adjusting/adding/removing text, tables or figures.

\section{Data}
\label{data}
Gaia Data Release 3 (DR3; \citealt{gaia23}) encompasses 33,812,183 sources, providing the measurements of radial velocities ($v_r$) and five-parameter astrometric data, including Right Ascension (R.A.; $\alpha$), Declination (decl.; $\delta$), parallaxes ($\varpi$), proper motions in R.A. ($\mu_{\alpha}$) and proper motions in decl. ($\mu_{\delta}$). Due to more observational data and improved data reduction pipeline, {\it Gaia} DR3 provides much more targets with radial velocity data than the 7,224,631 sources provided by {\it Gaia} Data Release 2 (DR2; \citealt{gaia18}), and the accuracy of radial velocity has also been significantly improved \cite[]{refId02}.

While distance is inherently a positive quantity, in some instances, sources may exhibit negative parallaxes due to various reasons. Bayesian inference has been adopted to estimate the distances for stars with negative parallax measurements \citep[e.g.,][]{2016ApJ...832..137A,refId0,2018AJ....156...58B,2021AJ....161..147B}. Following the recommendation of \cite{2015PASP..127..994B}, incorporating appropriate priors becomes crucial for {\it Gaia} sources with fractional parallax error ($\sigma_{\varpi}/\varpi$) exceeding 20\% for distance inference. 
However, targets with precise parallax measurements ($0 < \sigma_\varpi/\varpi < 0.2$) and high signal to noise ($S/N>5$) are not sensitive to priors \citep[e.g.,][]{2018ApJ...868...25B}.

Consequently, after correcting for zero-point offset in parallax \citep[][]{2021A&A...649A...4L, 2024A&A...691A..81D}, as well as magnitude and color-dependent proper motion bias \citep[][]{refId01}, we select a subset of 31,129,169 stars from {\it Gaia} DR3, which have valid radial velocity measurements and have fractional parallax error satisfying $0<\sigma_{\varpi}/\varpi<0.2~.$ This allows distance to be directly represented by the reciprocal of parallax, ensuring reliability and accuracy while minimizing potential bias from priors.

\section{Error propagation}
\label{method}
In this section, we delineate three error propagation methods utilized in this study. Our approach involves converting the five-parameter astrometry and radial velocities of {\it Gaia} sources to Cartesian coordinates and velocities in both the Equatorial and Galactic coordinate systems. Additional details regarding the transformation of galactic coordinates can be found in Appendix \ref{sec:e2g}.
%The transformation formulae are introduced in appendix \ref{sec:e2c} and \ref{sec:e2g}. 

\subsection{Linear error propagation}
Linear error propagation is the default method used by the most astronomical data analyses. Considering that linear error propagation is broadly used in the community \citep[e.g.][]{2014A&A...570A..62B}, we only briefly introduce the linear error propagation as follows.
%%change:In astrometric surveys like {\it Gaia} \citep{gaia18}, five-parameter solutions are typically provided for single stars, including R.A. ($\alpha$), delc. ($\delta$), parallax ($\varpi$), proper motion in the increasing R.A. direction ($\mu_\alpha$), and proper motion in the increasing decl. direction ($\mu_\delta$)
%%Gaia also measure the radial velocity ($v_r$) for tens of millions of stars.

First of all, we define the vectors and matrices that will be used in error propagation. The spherical ($\rm sph$) position and velocity vector in the equatorial coordinate system is defined as:
\begin{equation}
{\bm \xi}_{\rm sph}\equiv
\begin{bmatrix}
    \alpha\\
    \delta\\
    \varpi\\
    \mu_{\alpha^*}\\
    \mu_\delta\\
    v_r
\label{sph}
\end{bmatrix}~.
\end{equation}

The distance $r$ is derived from the parallax according to $r = \frac{\rm A}{\varpi}~$, where $\rm {A}=1\,\rm{AU}$. The position and velocity vector in Cartesian equatorial ($\rm eqt$) coordinates is linked to the spherical coordinates as: 

\begin{equation}
{\bm \xi}_{\rm eqt}\equiv
\begin{bmatrix}
    x\\
    y\\
    z\\
    v_x\\
    v_y\\
    v_z
\end{bmatrix}
    =
\begin{bmatrix}
    r\cos{\delta}\cos{\alpha}\\
    r\cos{\delta}\sin{\alpha}\\
    r\sin{\delta}\\
    v_{r}\cos{\delta}\cos{\alpha}-r\mu_{\alpha^*}\sin{\alpha}-r\mu_{\delta}\sin{\delta}\cos{\alpha}\\
    v_{r}\cos{\delta}\sin{\alpha}+r\mu_{\alpha^*}\cos{\alpha}-r\mu_{\delta}\sin{\delta}\sin{\alpha}\\
    r\mu_{\delta}\cos{\delta}+v_{r}\sin{\delta}
\end{bmatrix}~.
\label{eq:carx}
\end{equation}

%\begin{equation}
%    \Bar{F}=F(\Bar{q})
%	\label{eq:linbar}
%\end{equation}

%\begin{equation}
%    \sigma_{F}^2=\sum_{i=1}^{n} (\frac{\partial{F}}%{\partial{q_i}}\sigma_{q_i})^2 ~.
%	\label{eq:linsig}
%\end{equation}
The {\it Gaia} DR3 catalog provides the uncertainties, the corresponding correlation coefficients of the five-parameter astrometric solutions, and the errors of radial velocities. Considering that the radial velocity and the five-parameter astrometry are measured independently, we treat the five astrometric parameters as independent of the radial velocity. Hence the correlation coefficients between $v_{r}$ and the five astrometric parameters are zero. From {\it Gaia} DR3, we change $\sigma_{\alpha^*}$ to $\sigma_{\alpha}$, and use $\vec{\sigma}=(\sigma_{\alpha}, \sigma_{\delta}, \sigma_{\varpi}, \sigma_{\mu_{\alpha^*}}, \sigma_{\mu_\delta}, \sigma_{v_r})$ to denote the errors in the position and velocity vectors for sources, and $\rho_{ij}\,(=\rho_{ji})$ to denote the corresponding correlation coefficient, with i or j representing $\alpha$, $\delta$, $\varpi$, $\mu_{\alpha^*}$, $\mu_\delta$, or $v_r$. The six-dimensional variance-covariance matrix (or covariance matrix) is:

\begin{equation}
    \bm{C}_{\rm sph} = 
\begin{bmatrix}
    \sigma^2_{\alpha} & \sigma_{\alpha}\sigma_{\delta}\rho_{\alpha \delta} & \cdots & \sigma_{\alpha}\sigma_{\mu_{\delta}}\rho_{\alpha \mu_{\delta}} & 0\\
    \sigma_{\delta}\sigma_{\alpha}\rho_{\alpha \delta} & \sigma_{\delta}^2 & \cdots & \sigma_{\delta}\sigma_{\mu_{\delta}}\rho_{\delta \mu_{\delta}} & 0\\
    \vdots & \vdots  & \ddots   & \vdots & \vdots  \\
    \sigma_{\mu_{\delta}}\sigma_{\alpha}\rho_{\alpha \mu_{\delta}} & \sigma_{\mu_{\delta}}\sigma_{\delta}\rho_{\delta \mu_{\delta}} & \cdots & \sigma_{\mu_{\delta}}^2 & 0\\
    0 & 0 & \cdots & 0 & \sigma_{v_r}^2
\end{bmatrix}~.
\label{vc}
\end{equation}

The transformation from spherical coordinates to Cartesian coordinates requires the utilization of the Jacobian matrix:

\begin{equation}
    \bm{J} =
    \frac{\partial{(x, y, z, v_x, v_y, v_z)}}{\partial{(\alpha, \delta, \varpi, \mu_{\alpha^*}, \mu_\delta, v_r)}}
    =
\begin{bmatrix}
    \frac{\partial{x}}{\partial{\alpha}} & \frac{\partial{x}}{\partial{\delta}} & \cdots & \frac{\partial{x}}{\partial{v_{r}}}\\
    \frac{\partial{y}}{\partial{\alpha}} & \frac{\partial{y}}{\partial{\delta}} & \cdots & \frac{\partial{y}}{\partial{v_{r}}}\\
    \vdots & \vdots  & \ddots   & \vdots  \\
    \frac{\partial{v_z}}{\partial{\alpha}} & \frac{\partial{v_z}}{\partial{\delta}} & \cdots & \frac{\partial{v_z}}{\partial{v_{r}}}
\end{bmatrix}~,
\end{equation}
see Appendix \ref{j} for details of the Jacobian matrix. Therefore, the conversion from a covariance matrix in spherical coordinates to one in Cartesian equatorial coordinates can be expressed as:
\begin{equation}
    \bm{C}_{\rm eqt} = \bm{J}\bm{C}_{\rm sph}\bm{J}^{T}~.
\label{eq:ccar}
\end{equation}

\subsection{Second-order error propagation}
Although linear propagation is widely used, it may not always yield the best results for accurate error propagation \citep[e.g.][]{ilyin12}. Therefore, higher order error propagation is necessary to ensure optimal performance in certain cases. Specifically, when dealing with measurements that have relatively large errors, the truncation errors of the Taylor series can become significant. In such scenarios, the use of a linear transformation may result in decreased accuracy. To solve this problem and achieve enhanced precision, it is necessary to employ a second-order Taylor series for propagating statistical errors.

Second-order error propagation is commonly used in science and engineering, especially for precise calculations and data processing  \citep{wang2008nonparametric,10.1093/acprof:oso/9780198723844.003.0014}. In astronomy, second-order error propagation is frequently applied to measure the position and motion of  celestial bodies, such as satellite orbit calculations \citep[e.g.,][]{10.1007/s10569-006-9054-5,LI2020285}.

Following \cite{10.1115/1.1446068}, the second-order mean and variance of the output vector $\bm F$, transformed from those of the input vector $\bm b$, can be expressed as follows: 

\begin{equation}
    \bm{\Bar{F}}=\bm{F(\Bar{b})}+\frac{1}{2!}\sum_{j=1}^{n}\sum_{i=1}^{n} \frac{\partial^2{\bm{F}}}{\partial{b_i}\partial{b_j}}\sigma_{b_i}\sigma_{b_j} ~,
	\label{eq:secbar}
\end{equation}

\begin{equation}
    \sigma_{F}^2=\sum_{i=1}^{n} (\frac{\partial{\bm{F}}}{\partial{b_i}}\sigma_{b_i})^2+\frac{1}{2!}\sum_{j=1}^{n}\sum_{i=1}^{n} (\frac{\partial^2{\bm{F}}}{\partial{b_i}\partial{b_j}}\sigma_{b_i}\sigma_{b_j})^2 ~,
	\label{eq:secsig}
\end{equation}
where $\bm{F(\Bar{b})}$ and $\sum_{i=1}^{n} (\frac{\partial{\bm{F}}}{\partial{b_i}}\sigma_{b_i})^2$ are the first-order mean and variance.

When the Hessian matrix $\bm{H}$ and the covariance matrix $\bm{C}$ are available, calculations of the second-order moments about vector $\bm{F}$ and variables $b_i$, where i=1,...,n, can be performed using matrix operations (see \cite{5977574} for detailed definitions and derivation).

\iffalse
\begin{equation}
    \bm{H} = \bm{H(F)} =
\begin{bmatrix}
    \frac{\partial^2{F}}{\partial{b_1}^2} & \frac{\partial^2{F}}{\partial{b_1}\partial{b_2}} & \cdots & \frac{\partial^2{F}}{\partial{b_1}\partial{b_n}}\\
    \frac{\partial^2{F}}{\partial{b_2}\partial{b_1}} & \frac{\partial^2{F}}{\partial{b_2}^2} & \cdots & \frac{\partial^2{F}}{\partial{b_2}\partial{b_n}}\\
    \vdots & \vdots  & \ddots   & \vdots  \\
    \frac{\partial^2{F}}{\partial{b_n}\partial{b_1}} & \frac{\partial^2{F}}{\partial{b_n}\partial{b_2}} & \cdots & \frac{\partial^2{F}}{\partial{b_n^2}}
\end{bmatrix}
\end{equation}

\begin{equation}
    \bm{C} = \bm{C(b)} = 
\begin{bmatrix}
    \sigma_{b_1}^2 & \sigma_{b_1}\sigma_{b_2} & \cdots & \sigma_{b_1}\sigma_{b_n}\\
    \sigma_{b_2}\sigma_{b_1} & \sigma_{b_2}^2 & \cdots & \sigma_{b_2}\sigma_{b_n}\\
    \vdots & \vdots  & \ddots   & \vdots \\
    \sigma_{b_n}\sigma_{b_1} & \sigma_{b_n}\sigma_{b_2} & \cdots & \sigma_{b_n}^2

\end{bmatrix}
\end{equation}
\fi

In matrix form, the second-order terms in the aforementioned equations can be expressed as:

\begin{equation}
    \sum_{j=1}^{n}\sum_{i=1}^{n} \frac{\partial^2{\bm{F}}}{\partial{b_i}\partial{b_j}}\sigma_{b_i}\sigma_{b_j} = tr(\bm{H C})~,
\label{2mean}
\end{equation}

\begin{equation}
    \sum_{j=1}^{n}\sum_{i=1}^{n} (\frac{\partial^2{\bm{F}}}{\partial{b_i}\partial{b_j}}\sigma_{b_i}\sigma_{b_j})^2 = tr(\bm{H C H C})~.
\label{2var}
\end{equation}

Performing a Taylor series expansion at a particular point provides results that hold true only in the immediate proximity of that point. Consequently, the precision of the approximation tends to diminish as the deviation from the mean grows, particularly when dealing with a non-Gaussian distribution of the output vector. While it is feasible to analytically derive the higher moments of a non-Gaussian distribution, reconstructing the distribution itself uniquely from these higher moments poses a considerable challenge. Given that many astronomical applications primarily focus on the mean and covariance, we opt not to undertake error propagation for higher moments in this study.

\subsection{Monte Carlo (MC) error propagation}
To determine the precision of linear and second-order error propagation methods, we use the MC error propagation method as the reference for comparisons. Instead of directly sampling the absolute positions ($\alpha_0$, $\delta_0$), we sample the deviations relative to the observed positions according to the observational errors. In order to prevent a bias toward the equatorial poles, we generate samples of coordinates and velocities ($\Delta\alpha^*$, $\Delta\delta$, $\varpi$, $\mu_{\alpha^*}$, $\mu_\delta$, $v_r$) from 6-dimensional joint Gaussian distributions centered on the observed values of (0 ,0 , $\varpi_0$, $\mu_{\alpha^*0}$, $\mu_{\delta0}$, $v_{r0}$). The covariance matrix is defined using the standard derivations ($\sigma_{\alpha^*}, \sigma_{\delta}, \sigma_{\varpi}, \sigma_{\mu_{\alpha^*}}, \sigma_{\mu_\delta}, \sigma_{v_r}$). To obtain samples of spherical positions, we add the sampled position deviations to the observed positions: 
\begin{equation}
    \alpha = \frac{\Delta\alpha^*}{\cos{\delta_0}} + \alpha_0 ~,\\
\end{equation}
\begin{equation}
    \delta = \Delta\delta + \delta_0~.
\end{equation}
 %In order to compare linear and second-order propagation techniques, 10 million MC samples (named MC7) are drawn from the covariance of {\it Gaia} astrometry. 
 This aims to avoid rounding errors, because the corresponding observational errors ($\sigma_{\alpha}, \sigma_\delta$) are very small compared with $\alpha_0$ and $\delta_0$ themselves.

 For {\it Gaia} DR3 with radial velocity and fractional parallax error satisfying $0<\sigma_{\varpi}/\varpi<0.2$, only about 0.001\% of the data show negative parallaxes in MC samples. We resample until we achieve the MC sample with positive parallaxes. This resampling has minimal impact on the mean and covariance, as the fractional bias caused by this process is approximately 0.01\%. We calculate the Cartesian coordinates for each MC sample according to Eq.~\ref{eq:carx}, and get the distribution of Cartesian coordinates.

To more accurately evaluate the error propagation methods, we generate a reference set of 10 million Monte Carlo samples (referred to as MC7) for comparison. Additionally, we employ Monte Carlo simulations with $10^3$ (MC3), $10^4$ (MC4), $10^5$ (MC5), and $10^6$ (MC6) samples. Transforming each sample from spherical to Cartesian coordinates in the Equatorial system and using their means as the MC propagation results, we then define the fractional deviations of coordinates and velocities ($\frac{\Delta r}{r}, \frac{\Delta v}{v}$) and the fractional deviations of their variances ($\frac{\sigma_{\Delta_r}}{\sigma_{r}}, \frac{\sigma_{\Delta_v}}{\sigma_{v}}$) relative to MC7 using various methods:

\begin{equation}
    \frac{\Delta r}{r}=\frac{\sqrt{(x-x_{\rm MC7})^2+(y-y_{\rm MC7})^2+(z-z_{\rm MC7})^2}}{\sqrt{x_{\rm MC7}^2+y_{\rm MC7}^2+z_{\rm MC7}^2}}~,
	\label{eq:dr}
\end{equation}

\begin{equation}
    \frac{\Delta v}{v}=\frac{\sqrt{(v_x-v_{x,{\rm MC7}})^2+(v_y-v_{y,{\rm MC7}})^2+(v_z-v_{z,{\rm MC7}})^2}}{\sqrt{v_{x,{\rm MC7}}^2+v_{y,{\rm MC7}}^2+v_{z,{\rm MC7}}^2}}~,
	\label{eq:dv}
\end{equation}

\begin{equation}
    \frac{\sigma_{\Delta_r}}{\sigma_{r}}=\frac{\sqrt{(\sigma_{x}-\sigma_{x,{\rm MC7}})^2+(\sigma_{y}-\sigma_{y,{\rm MC7}})^2+(\sigma_{z}-\sigma_{z,{\rm MC7}})^2}}{\sqrt{\sigma_{x,{\rm MC7}}^2+\sigma_{y,{\rm MC7}}^2+\sigma_{z,{\rm MC7}}^2}}~,
	\label{eq:dsigr}
\end{equation}

\begin{equation}
    \frac{\sigma_{\Delta_v}}{\sigma_{v}}=\frac{\sqrt{(\sigma_{v_x}-\sigma_{v_{x,{\rm MC7}}})^2+(\sigma_{v_y}-\sigma_{v_{y,{\rm MC7}}})^2+(\sigma_{v_z}-\sigma_{v_{z,{\rm MC7}}})^2}}{\sqrt{\sigma_{v_{x,{\rm MC7}}}^2+\sigma_{v_{y,{\rm MC7}}}^2+\sigma_{v_{z,{\rm MC7}}}^2}} ~.
	\label{eq:dsigv}
\end{equation}

\section{Results}
\label{results}

\begin{figure*}
	\centering
	\begin{minipage}{0.49\linewidth}
		\centering
		\includegraphics[width=1\linewidth]{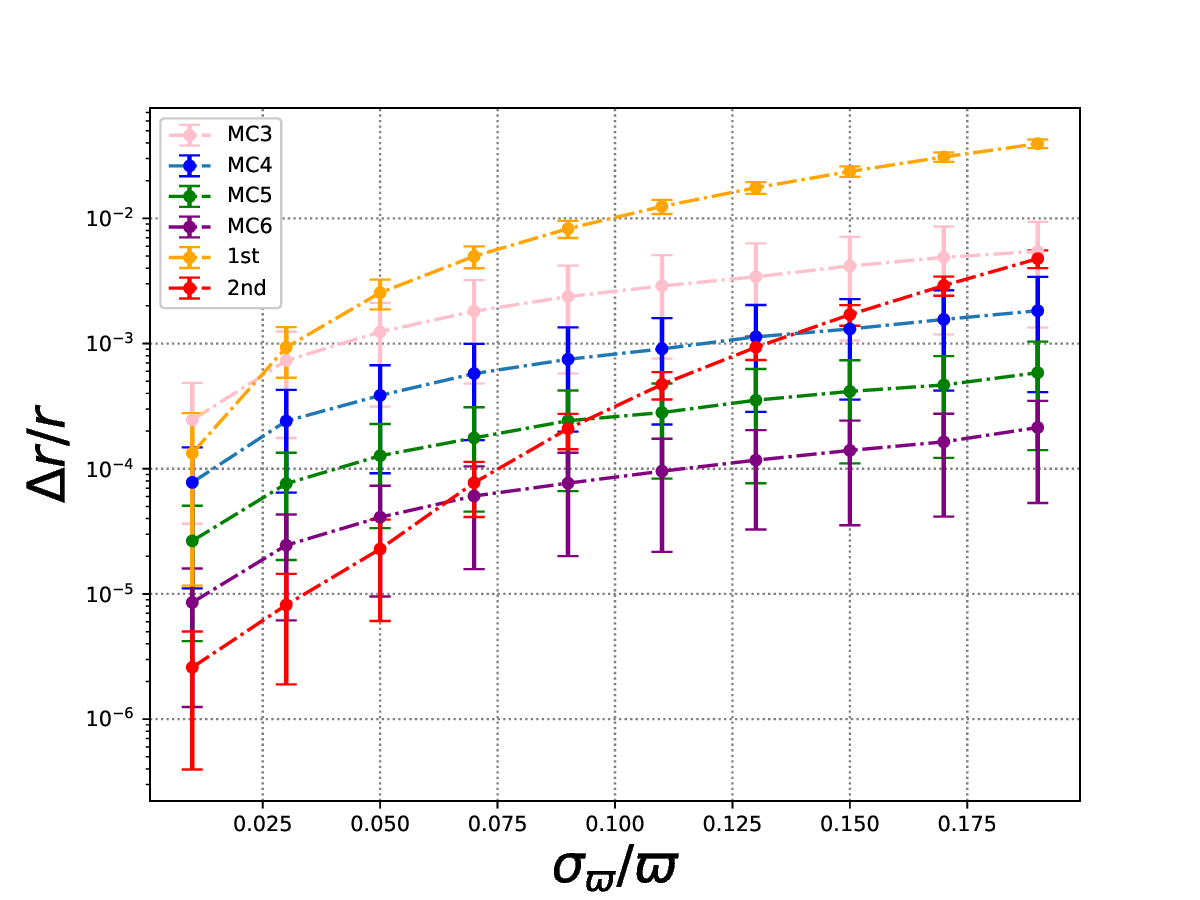}
	\end{minipage}
	\begin{minipage}{0.49\linewidth}
		\centering
		\includegraphics[width=1\linewidth]{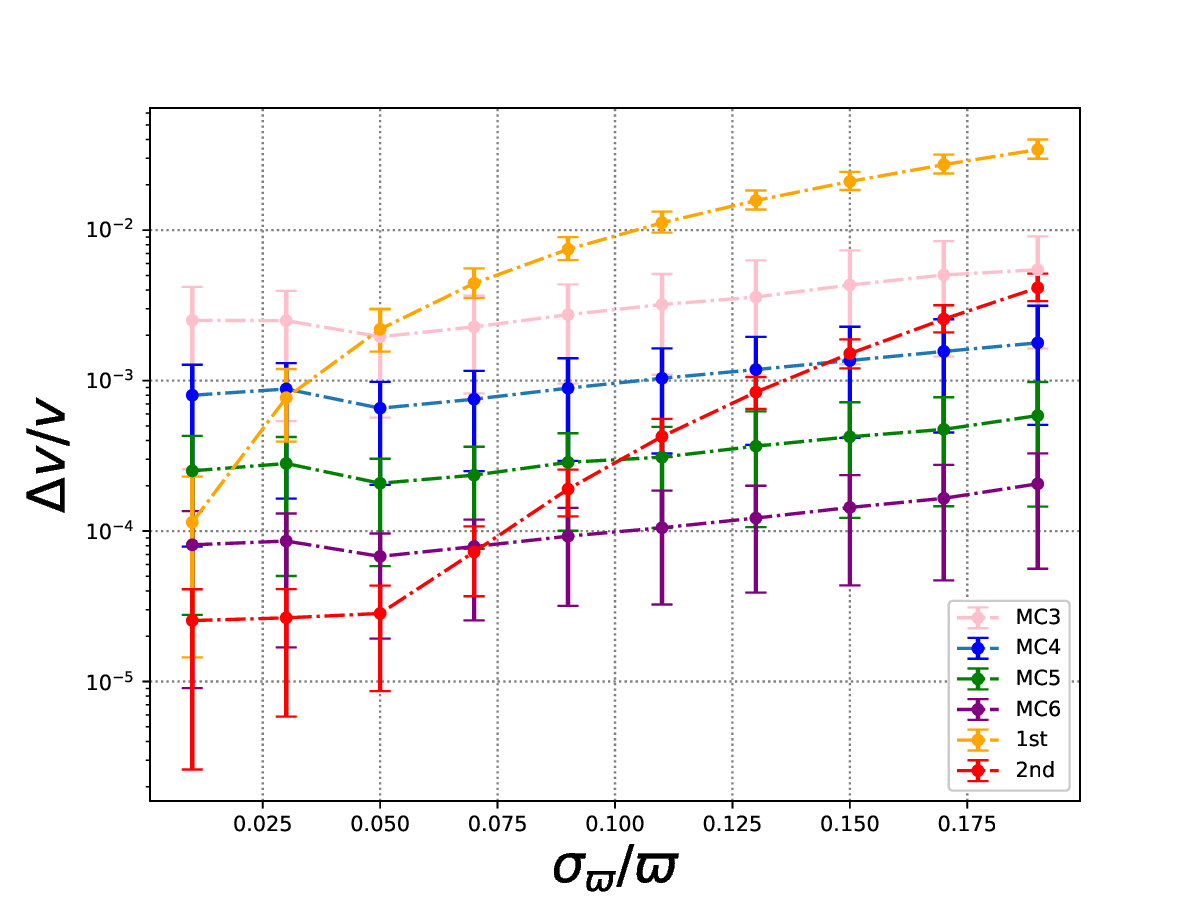}
	\end{minipage}
	%\qquad
	%让图片换行，
	
	\begin{minipage}{0.49\linewidth}
		\centering
		\includegraphics[width=1\linewidth]{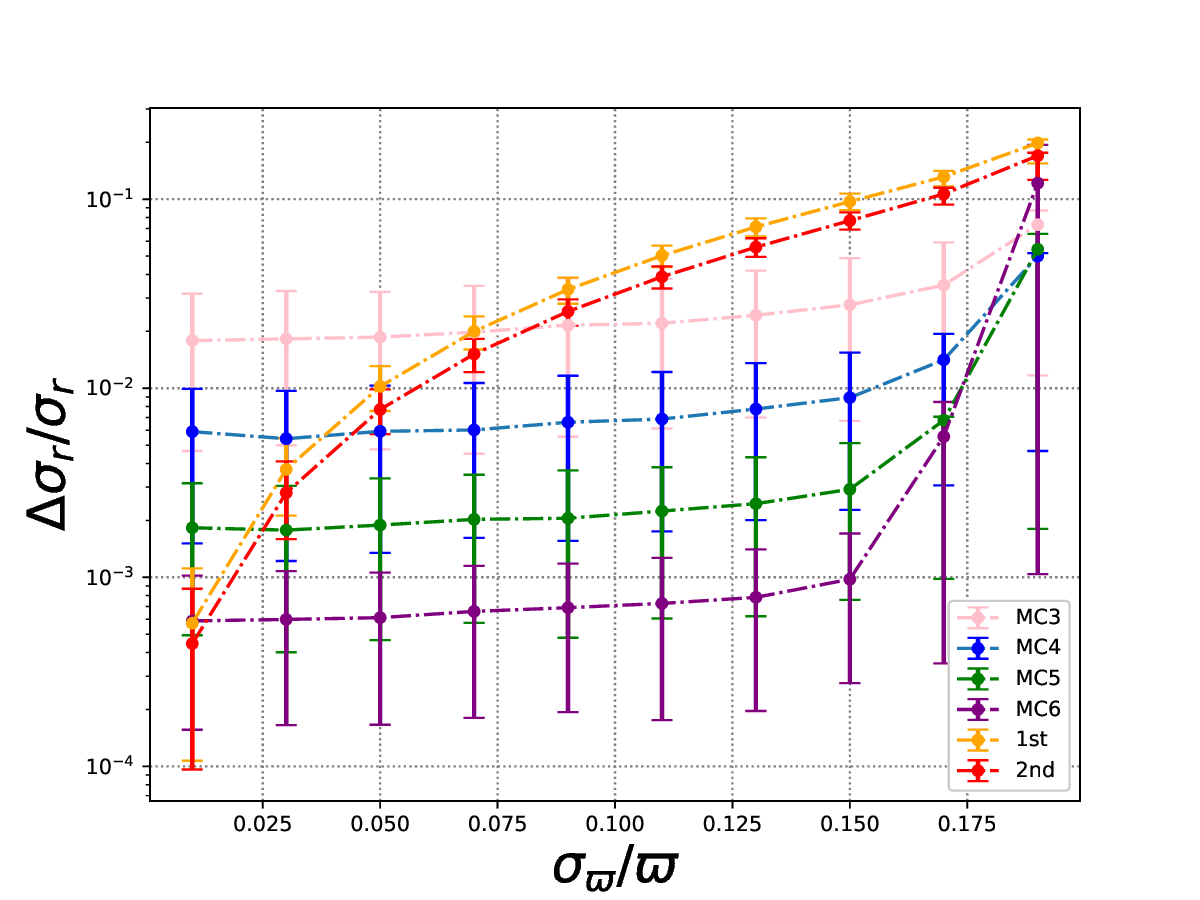}
	\end{minipage}
	\begin{minipage}{0.49\linewidth}
		\centering
		\includegraphics[width=1\linewidth]{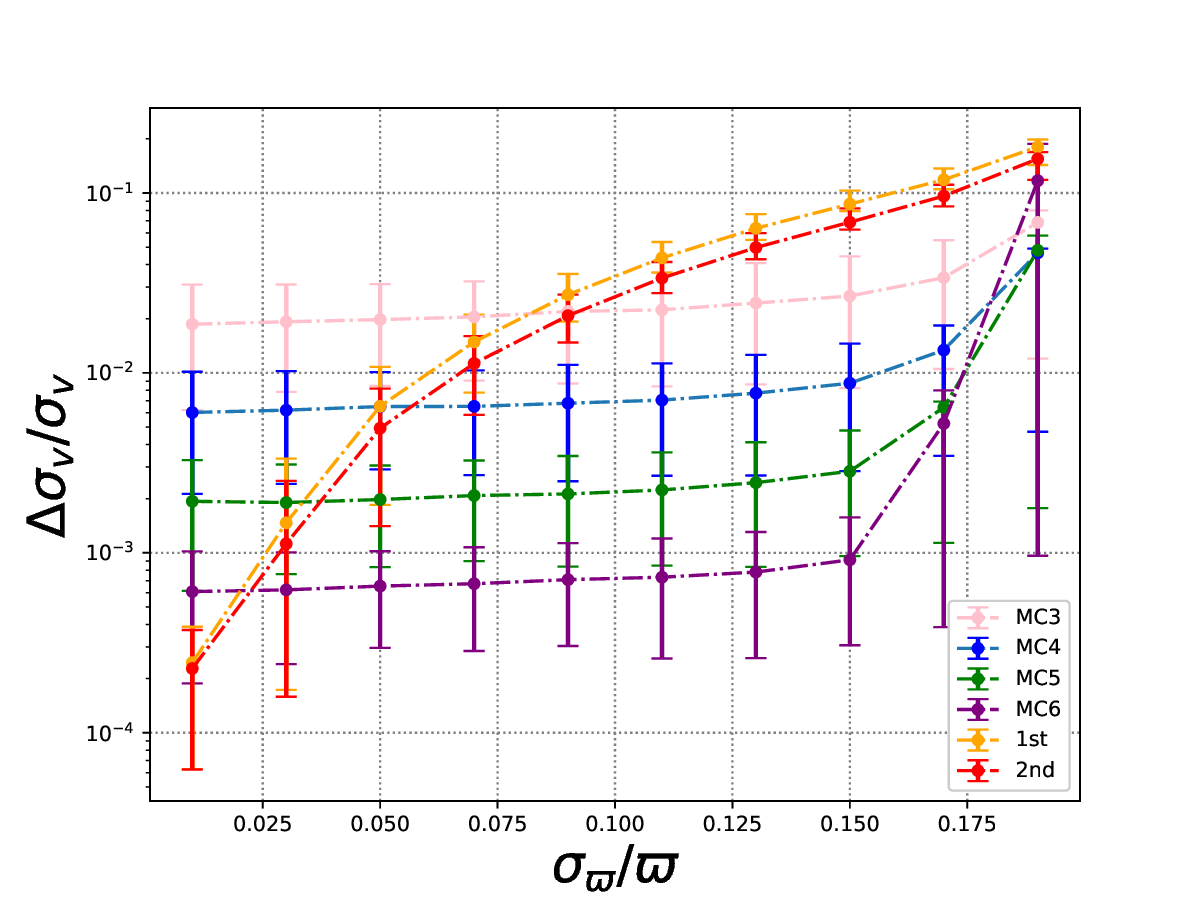}
	\end{minipage}
    \caption{Comparison of first-order ($\rm {1st}$), second-order ($\rm {2nd}$), and MC error propagation methods. Top panels: the fractional deviations of coordinate (left) and velocity (right) propagated with various methods relative to MC7. Bottom panels: the fractional deviations of the variances of coordinate (left) and velocity (right) propagated with various methods relative to $\rm MC7$. The x axis shows the fractional parallax error ($\sigma_\varpi/\varpi$) and 10 bins are used to divide the range of $0<\sigma_\varpi / \varpi<0.2$. Each dot and error bar respectively represent the average and scatter of the values of 1000 samples. The error bars correspond to quantiles of 16\% and 84\%. }
	\label{fig:drdv}%文中引用该图片代号
\end{figure*}

To investigate the dependence of error propagation on the fractional parallax error, we evenly divide $\sigma_\varpi/\varpi$ (with a range of (0, 0.2)) into 10 bins and randomly select 1000 stars from {\it Gaia} DR3 for each bin. We employ the linear, second-order, and MC error propagation methods with different sample sizes to transform stars from spherical coordinates and velocities to Cartesian coordinates and velocities in the Equatorial system. We compare the fractional deviations of the mean and variance of various methods relative to MC7 in Fig.~\ref{fig:drdv}.

By analyzing the results, we find the following features:
\begin{itemize}
\item {\bf Propagation of coordinate and velocity:}
According to the top panels of Fig. \ref{fig:drdv}, the linear error propagation will lead to larger than 1\% fractional deviations of mean coordinate and velocity for targets with $\sigma_\varpi/\varpi>0.11$. The propagation of mean coordinate and velocity using the second-order error propagation and the MC propagation method with more than 10,000 samples achieves less than 0.2\% precision for sources with $0<\sigma_\varpi/\varpi<0.15$. The second-order error propagation achieves less than 0.5\% precision for sources with $0<\sigma_\varpi/\varpi<0.2$. 

\item {\bf Propagation of the variance: }
As seen from the bottom panels of Fig. \ref{fig:drdv}, the linear and second-order variance propagation can achieve 10\% precision for sources with less than 15\% fractional parallax error. In contrast, MC method with more than 10,000 samples propagates variance with less than 1\% precision for sources with $0<\sigma_{\varpi}/{\varpi}<0.15$. 
While the second-order error propagation can achieve higher precision for mean transformation than the MC4 method for sources with small fractional parallax error, it fails to propagate variance as precise as MC4. The reason for this is that Gaussian distributions in the spherical coordinate systems are transformed into non-Gaussian distributions in the Cartesian coordinate system. Because the higher moments are not taken into account in calculating the variance of non-Gaussian distributions, the second-order error propagation is less accurate in propagating the variance (and covariance) than propagating the mean. 
%Hence we recommend the second-order method for $10$\%-precision variance propagation while recommend MC4 for $1$\%-precision variance propagation. This also applies to the covariance propagation. 
\end{itemize}

\begin{figure}
\centering
\includegraphics[width=0.8\linewidth]{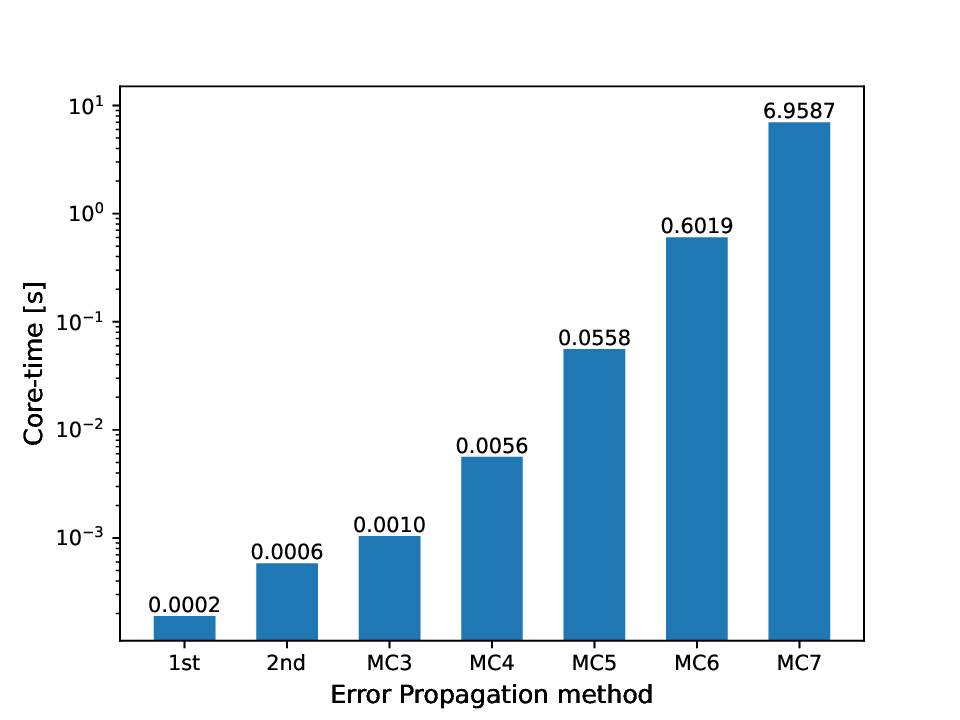}
\caption{Comparison of the CPU time of the various propagation methods. Each bar in the chart shows a single run time for the equatorial Cartesian coordinate and covariance transformation of the respective propagation method, which is taken as an average of 10,000 runs. The CPU model is \emph{Intel(R) Xeon(R) Silver 4210R CPU @ 2.40GHz }. }
\label{fig:time}
\end{figure}

To evaluate the efficiency of various propagation methods, we calculate the computational time of mean and covariance for each method in Fig.~\ref{fig:time}. The time consumed by MC method is linearly proportional to the number of MC samples. Meanwhile the execution time for second-order error propagation is one order of magnitude lower than that of MC4. Although MC3 requires a comparable amount of time as second-order error propagation, the latter exhibits significantly higher precision in propagation, as illustrated in Fig.~\ref{fig:drdv}. 

Therefore, considering the efficiency and precision of error propagation, we recommend the second-order error propagation for mean transformation with approximately 0.5\% precision and for covariance propagation with 10\% precision. If higher precision of variance propagation is needed, we recommend using the MC4 method. Combining the advantages of both methods, we apply the second-order error propagation to convert the means, and employ MC4 to propagate the covariance of astrometric parameters and radial velocities of 30 million {\it Gaia} sources with $0<\sigma_{\varpi}/{\varpi}<0.2$ into Cartesian coordinates and velocities in both the Equatorial and Galactic coordinate systems. In this way, the average CPU time required for a single calculation of a set of Cartesian catalog data (including both galactic and equatorial coordinate systems) is only 0.0065\,s. 

\begin{figure*}
	\centering
        \begin{minipage}{0.49\linewidth}
		\centering
		\includegraphics[width=1\linewidth]{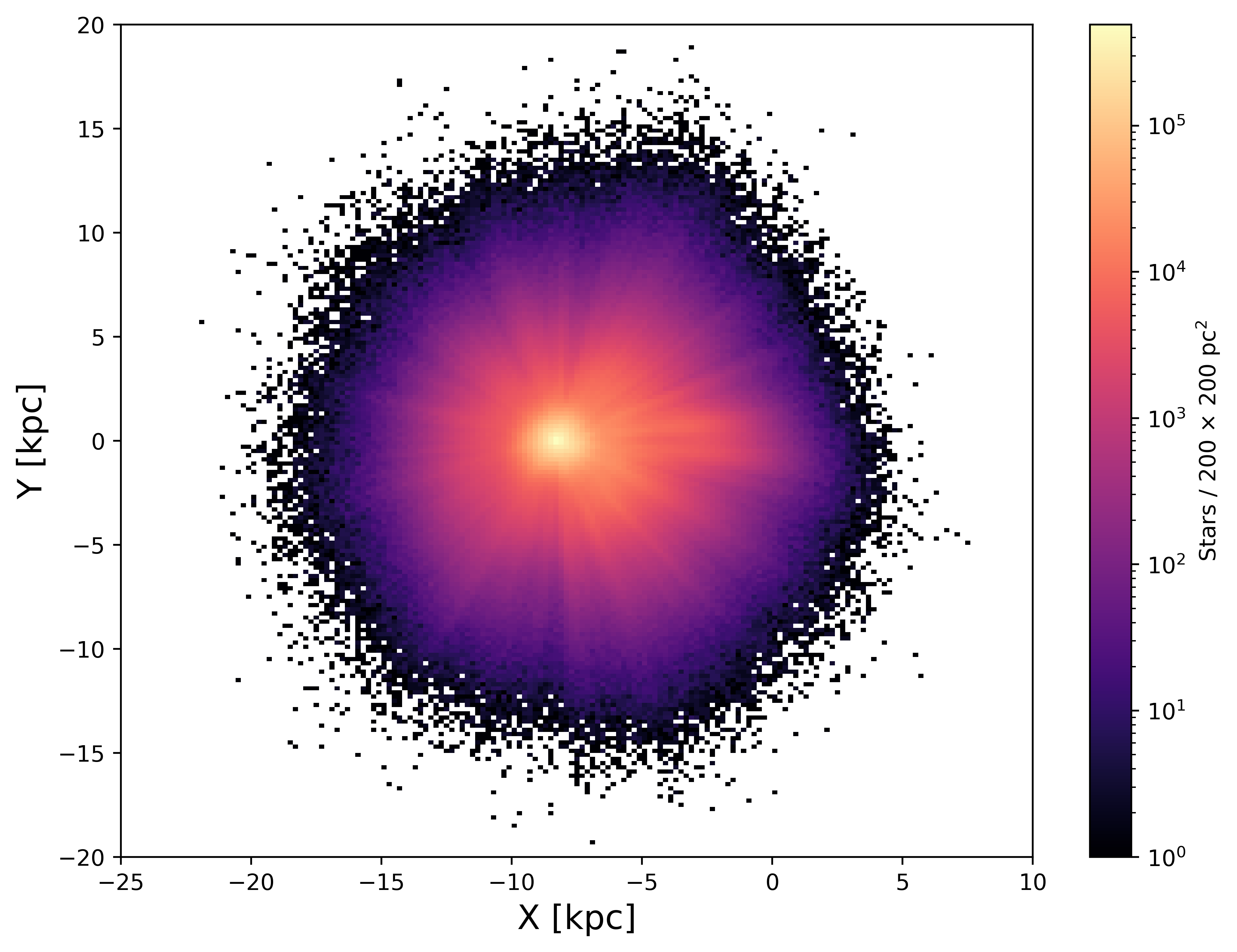}
	\end{minipage}
	\begin{minipage}{0.49\linewidth}
		\centering
		\includegraphics[width=1\linewidth]{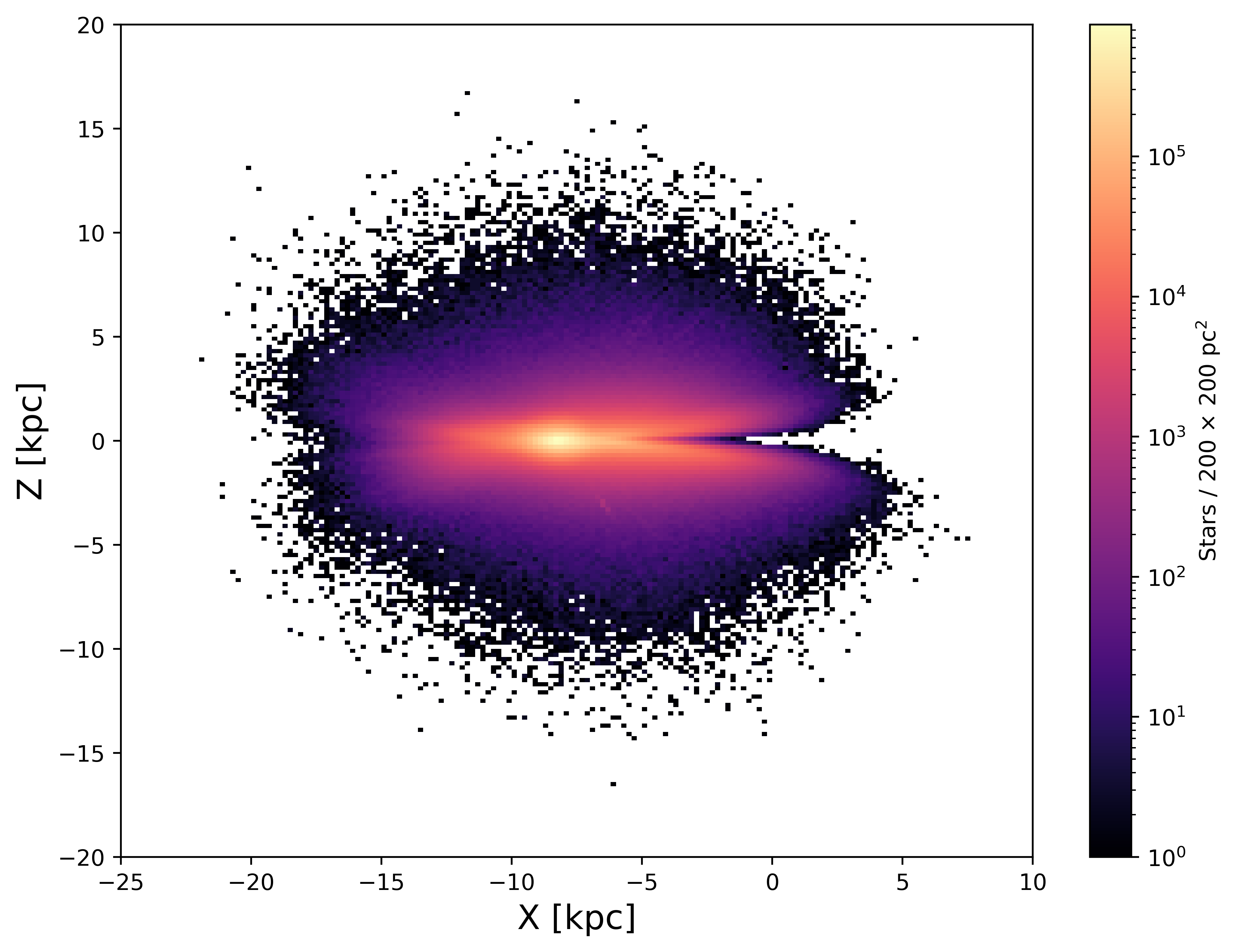}
	\end{minipage}

    \caption{Spatial distributions of stars from our Cartesian catalog (similar to Fig. 2 in \cite{refId02}). Left: a face-on view of catalog stars in galactocentric Cartesian coordinates (X, Y). Right: an edge-on view of the same data in galactocentric Cartesian coordinates (X, Z). The Galactic center is at X = Y = 0 kpc, with the Sun positioned at X = -8.277 kpc \cite[]{2022A&A...657L..12G}, Y = 0 kpc, and Z = 20.8 pc \cite[]{10.1093/mnras/sty2813}. The Milky Way is viewed from the Galactic north pole and rotates clockwise, with the Z-axis positive towards the north pole. The stars are binned by a \rm{200 pc $\times$ 200 pc} grid. }
	\label{fig:xyz}
\end{figure*}

Fig.~\ref{fig:xyz} shows the galactocentric Cartesian coordinate distribution of stars. The figure displays a sample of 31,066,855 stars from {\it Gaia} DR3 with radial velocity, where the parallax satisfies $0<\sigma_{\varpi}/\varpi<0.2$ and duplicated sources are removed. Although the face-on view (XY plane) is similar to Fig.\,2 in \cite{refId02}, Fig.~\ref{fig:xyz} doesn't have elongated features in face-on and edge-on views, because the catalog doesn't contain Large and Small Magellanic Cloud stars due to their fractional parallax errors larger than 20\% (as discussed in \cite{2021AJ....161..147B}).  

\begin{figure}
\centering
\includegraphics[width=0.8\linewidth]{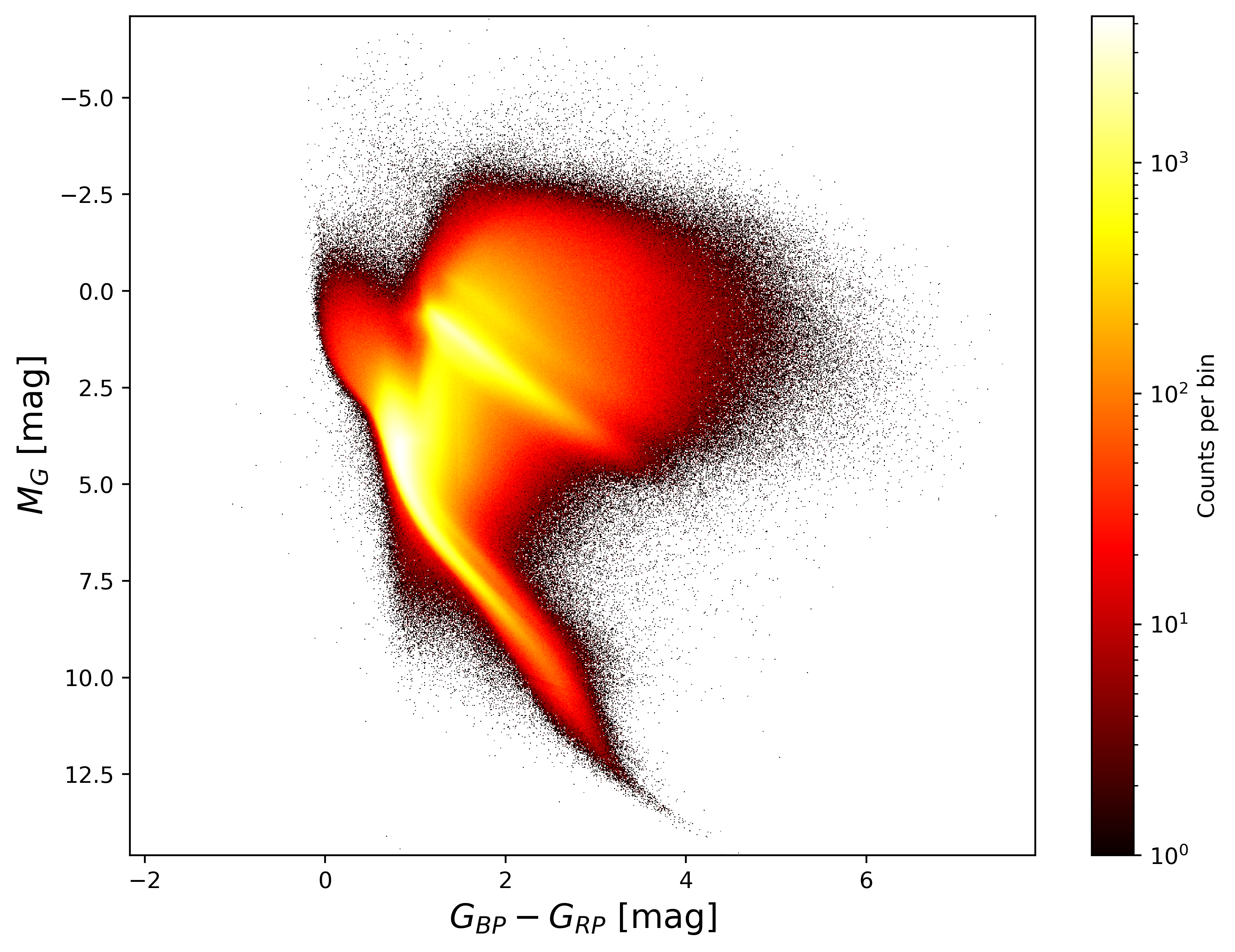}
\caption{Hertzsprung-Russell diagram of the Cartesian catalog. The colorbar represents the frequency of stars in {\it Gaia} DR3 with radial velocity and fractional parallax error satisfying $0<\sigma_{\varpi}/\varpi<0.2$. }

\label{fig:hr}
\end{figure}

The Hertzsprung-Russell (HR) diagram of the Cartesian catalog is shown in Fig.~\ref{fig:hr}, which includes only data with $M_G$ and $G_{BP}-G_{RP}$. The figure shows that the Cartesian catalog contains stars with absolute magnitudes ranging from 15 to -5, including a large number of red giants and main-sequence stars, but no white dwarfs. The spectroscopic pipeline of {\it Gaia} lacks an appropriate template for white dwarfs, and the mismatch between the observed spectra and the templates can lead to significant systematic errors in radial velocity measurements \citep[][]{refId02}. Consequently, white dwarfs lack radial velocity data and are not included in this catalog. 

The Cartesian catalog in Appendix~\ref{apd:data} contains the mean of Cartesian coordinates determined by the second-order error propagation and the covariance of Cartesian coordinates determined by the MC4 method. We also include the values of transversal velocity and distance in the catalog. 

\section{Discussion and conclusion}
\label{conclusion}

In our study based on {\it Gaia} DR3 data, we compare various error propagation methods and identify second-order error propagation as the most efficient for achieving nearly 0.5\% precision in propagating mean coordinates and velocities. Linear propagation commonly used fails to achieve 1\% precision in coordinate and velocity mean when the fractional parallax error exceeds 10\%.

However, the precision of second-order error propagation for variance is inferior to that for mean transformation. This discrepancy arises due to the nonlinear coordinate transformation introducing non-Gaussian errors in new coordinate systems. 
Although the MC4 method is about 10 times more computationally expensive than the second-order method, it achieves more accurate covariance propagation than the latter. Therefore, for propagating covariance with nearly 1\% precision, we suggest employing the MC propagation method with a minimum of 10,000 samples.

Balancing efficiency and precision, we employ second-order error propagation for the mean values of 31,129,169 {\it Gaia} sources with radial velocity and fractional parallax error below 20\%, and utilize MC4 for covariance conversion. We present the Cartesian coordinates, velocities, and their covariances in both Equatorial and Galactic coordinate systems. This catalog offers highly precise mean coordinates and velocities, facilitating applications such as the accurate integration of stellar orbits and the study of wide binaries.

Subsequent investigations into precise error propagation should address the non-Gaussian nature of errors and incorporate higher moments, including skewness and kurtosis, in the propagation process. Advanced techniques such as the Kalman and unscented filters (mentioned in Section \ref{sect:intro}) may offer more efficient and accurate results for error propagation in stellar motions. 
%% change: delete \citep[e.g.,][]{smith62, schmidt66, julier04, Chen2017}, repeat with penultimate paragraph of Introduction. 
%% change: error propagation methods-->error propagation results

\section*{Data availability}

The complete data in Table \ref{table:T1} will be available at CDS via anonymous ftp to cdsarc.u-strasbg.fr (130.79.128.5) or via https://cdsarc.cds.unistra.fr/viz-bin/cat/I/363 and the source code can be found on GitHub (https://github.com/LuyaoZhang-sjtu/GaiaDR3-error-propagation).

\begin{acknowledgements}
We sincerely thank the anonymous referees of RAA for their valuable comments, which significantly enhanced the quality of this paper.

This work is supported by the National Key R\&D Program of China, No. 2024YFA1611801 and No. 2024YFC2207800, by the National Natural Science Foundation of China (NSFC) under Grant No. 12473066.This work is supported by Shanghai Jiao Tong University 2030 Initiative. W.W. is supported by NSFC (Grant Nos. 12022307 and 12273021).

This work is based on data from the European Space Agency (ESA) mission {\it Gaia} (https://www.cosmos.esa.int/gaia), processed by the {\it Gaia} Data Processing and Analysis Consortium (DPAC, https://www.cosmos.esa.int/web/gaia/dpac/consortium).

\end{acknowledgements}

\appendix                  %%appendicial material is supported

\section{Data sample}
\label{apd:data} 
This appendix presents a sample of the data catalog in Table \ref{table:T1}, including the mean Cartesian coordinates derived from second-order error propagation, the covariance matrix obtained using the MC4 method, and the transversal velocity and distance for each source.

\begin{table*}
%\centering
\caption{Examples of the catalog of coordinates and velocities in Equatorial and Galactic coordinate systems based on second-order error propagation and their covariances based on MC4. The correlation coefficients between Cartesian coordinates are explained in section \ref{kij}. }
\label{table:T1}
\resizebox{\textwidth}{!}{
\begin{tabular}{ccccccccccc}
\toprule
       {\it Gaia} DR3 source\_id &     r\_eqt &  r\_eqt\_error &      vt\_eqt &  vt\_eqt\_error &     x\_eqt &  x\_eqt\_error &       x\_gal &  x\_gal\_error &     y\_eqt &  y\_eqt\_error \\

                  & [$\, \rm{pc}$] &  [$\, \rm{pc}$] &        [$\, \rm{km} \, \rm{s}^{-1}$] &  [$\, \rm{km} \, \rm{s}^{-1}$] &         [$\, \rm{pc}$] &  [$\, \rm{pc}$] &       [$\, \rm{pc}$] &  [$\, \rm{pc}$] &        [$\, \rm{pc}$]&        [$\, \rm{pc}$] \\

\midrule
200030787485939968 &  647.031553 &     8.665063 & 20601.543326 &    287.025171 &  134.532389 &     1.801530 &  -624.933163 &     8.368510 &  475.844692 &     6.372059 \\
243131265338696960 & 1467.466212 &    44.418498 & 41853.421674 &   1250.012167 &  627.700148 &    18.996902 & -1231.688678 &    37.276189 &  773.748036 &    23.416938 \\
171961316481695872 & 2046.674971 &    86.670104 & 10185.771205 &    462.796237 &  594.034343 &    25.149937 & -1978.215127 &    83.752711 & 1606.417789 &    68.011736 \\
 55616669682429568 &  255.495582 &    15.084233 & 11847.245619 &    815.448020 &  159.844142 &     9.448656 &  -206.101795 &    12.183024 &  183.827943 &    10.866379 \\
159237767925095808 & 4389.670253 &   396.858380 & 14210.820602 &   1362.759362 & 1334.125320 &   120.634047 & -4240.114346 &   383.398880 & 3524.523748 &   318.693871 \\
228161685104953472 & 3157.411196 &   351.706897 & 41874.578692 &   4738.891542 &  937.323592 &   104.518807 & -2955.617920 &   329.574186 & 2165.808759 &   241.504375 \\
176758313916240384 & 3546.819656 &   457.715455 & 34882.973858 &   4490.540202 & 1146.974741 &   148.282909 & -3352.638332 &   433.434972 & 2589.499195 &   334.775004 \\
110527723485651328 & 5401.092771 &   817.497730 & 62859.597739 &   9489.374298 & 3252.277495 &   493.056627 & -4472.803410 &   678.092618 & 3735.441241 &   566.305938 \\
266107794183243392 & 3948.204353 &   707.938159 & 30907.151439 &   5587.757422 &  539.050640 &    97.015368 & -3559.286905 &   640.580872 & 2298.168517 &   413.611724 \\
199834524659602560 & 6300.091255 &  1130.725447 & 47534.769063 &   8538.895743 & 1533.934200 &   276.429692 & -6018.874085 &  1084.658983 & 4569.422215 &   823.453820 \\
\bottomrule
\end{tabular}
}

\resizebox{\textwidth}{!}{
\begin{tabular}{cccccccccccc}
\toprule
    y\_gal &  y\_gal\_error &   z\_eqt &  z\_eqt\_error &       z\_gal &  z\_gal\_error &    vx\_eqt &  vx\_eqt\_error &     vx\_gal &  vx\_gal\_error &    vy\_eqt &  vy\_eqt\_error \\

  [$\, \rm{pc}$] &        [$\, \rm{pc}$] &  [$\, \rm{pc}$] &     [$\, \rm{pc}$] &  [$\, \rm{pc}$] &     [$\, \rm{pc}$] &  [$\, \rm{km} \, \rm{s}^{-1}$] &     [$\, \rm{km} \, \rm{s}^{-1}$] &  [$\, \rm{km} \, \rm{s}^{-1}$] &     [$\, \rm{km} \, \rm{s}^{-1}$] &  [$\, \rm{km} \, \rm{s}^{-1}$] & [$\, \rm{km} \, \rm{s}^{-1}$] \\
\midrule
 166.559433 &     2.230405 &  417.353424 &     5.588799 &   -20.676366 &     0.276878 & -15.519831 &      1.084849 &  55.609456 &      5.022529 & -32.056864 &      3.828528 \\
 770.979814 &    23.333160 & 1077.687511 &    32.615452 &  -206.487278 &     6.249192 & -12.399473 &      3.329478 &  -7.575050 &      6.562881 &  27.484758 &      4.250384 \\
 416.508218 &    17.633923 & 1121.275231 &    47.472006 &  -322.333480 &    13.646798 & -14.309175 &      0.556296 &  14.814779 &      1.215064 &  -9.772316 &      1.004909 \\
  53.973826 &     3.190484 &   75.993056 &     4.492078 &  -140.451711 &     8.302337 &  28.158682 &      3.147573 & -20.936957 &      4.014908 &  15.095567 &      3.607896 \\
 771.830412 &    69.790315 & 2249.637881 &   203.416364 &  -829.901657 &    75.041223 &  -2.715690 &      0.974283 & -23.395144 &      0.424860 &  25.014346 &      0.805159 \\
1062.875886 &   118.518856 & 2092.622279 &   233.343518 &  -288.077519 &    32.122864 &  -8.070650 &      2.798039 &  -1.796631 &      8.596169 &  22.500453 &      7.092617 \\
1001.847367 &   129.520587 & 2124.552047 &   274.665820 &  -539.348314 &    69.727897 &  -2.250695 &      2.588164 &  51.817585 &      3.789200 & -25.652130 &      3.466915 \\
1538.102217 &   233.181668 & 2132.255005 &   323.257306 & -2589.519802 &   392.580246 & -32.000688 &      5.176726 &  44.659904 &      7.084487 &  -6.910449 &      7.502853 \\
1594.456714 &   286.961546 & 3146.528012 &   566.294797 &   511.837053 &    92.117616 & -21.123758 &      3.874245 & -12.776766 &      8.054188 &  20.769009 &      6.272847 \\
1725.990732 &   311.040126 & 4017.062578 &   723.913302 &  -404.321664 &    72.862652 & -28.188661 &      6.384590 & -41.533855 &      3.926764 &  48.115034 &      5.407343 \\
\bottomrule
\end{tabular}

}

\resizebox{\textwidth}{!}{
\begin{tabular}{cccccccccccc}
\toprule
    vy\_gal &  vy\_gal\_error &     vz\_eqt &  vz\_eqt\_error &     vz\_gal &  vz\_gal\_error &  x\_y\_eqt\_corr &  x\_y\_gal\_corr &  x\_z\_eqt\_corr &  x\_z\_gal\_corr &  x\_vx\_eqt\_corr &  x\_vx\_gal\_corr \\

 [$\, \rm{km} \, \rm{s}^{-1}$] &     [$\, \rm{km} \, \rm{s}^{-1}$] &  [$\, \rm{km} \, \rm{s}^{-1}$] & [$\, \rm{km} \, \rm{s}^{-1}$] & [$\, \rm{km} \, \rm{s}^{-1}$] & [$\, \rm{km} \, \rm{s}^{-1}$] &   &   &    &    &  & \\
\midrule
-34.719950 &      1.363982 & -55.304265 &      3.361144 &  -5.402112 &      0.213469 &           1.0 &            -1 &           1.0 &           1.0 &      -0.025778 &       0.007938 \\
-42.669929 &      4.195569 & -32.553917 &      5.735937 &  -9.529537 &      1.173857 &           1.0 &            -1 &           1.0 &           1.0 &      -0.029167 &       0.108153 \\
-11.205423 &      0.432668 & -11.355213 &      0.701772 &   9.173458 &      0.384982 &           1.0 &            -1 &           1.0 &           1.0 &      -0.613767 &       0.089652 \\
 16.780904 &      1.228443 &  12.828138 &      1.502041 & -21.572986 &      2.748034 &           1.0 &            -1 &           1.0 &           1.0 &       0.167898 &      -0.092520 \\
 -9.850999 &      1.288472 &   3.504734 &      0.731835 &  -1.000336 &      0.540765 &           1.0 &            -1 &           1.0 &           1.0 &      -0.824537 &       0.634052 \\
-40.880532 &      5.207441 & -35.989984 &      6.824120 & -13.865027 &      1.921271 &           1.0 &            -1 &           1.0 &           1.0 &      -0.201012 &       0.159771 \\
-34.919209 &      2.647298 & -60.534173 &      4.225765 & -20.568671 &      3.899867 &           1.0 &            -1 &           1.0 &           1.0 &       0.819537 &       0.015970 \\
-69.657557 &      8.532614 & -76.199526 &      8.954340 &  -5.611040 &      6.346341 &           1.0 &            -1 &           1.0 &           1.0 &       0.026984 &       0.004039 \\
-26.167317 &      6.096840 &  -8.689895 &      7.205121 &  10.252069 &      2.056878 &           1.0 &            -1 &           1.0 &          -1.0 &      -0.922549 &       0.252909 \\
-33.702068 &      7.640647 &   2.181059 &      3.766084 &  15.922425 &      3.222852 &           1.0 &            -1 &           1.0 &           1.0 &      -0.981286 &       0.608243 \\
\bottomrule
\end{tabular}
}

\resizebox{\textwidth}{!}{
\begin{tabular}{cccccccccccc}
\toprule
 x\_vy\_eqt\_corr &  x\_vy\_gal\_corr &  x\_vz\_eqt\_corr &  x\_vz\_gal\_corr &  y\_z\_eqt\_corr &  y\_z\_gal\_corr &  y\_vx\_eqt\_corr &  y\_vx\_gal\_corr &  y\_vy\_eqt\_corr &  y\_vy\_gal\_corr &  y\_vz\_eqt\_corr &  y\_vz\_gal\_corr \\
\midrule
      0.044574 &       0.187635 &      -0.066197 &       0.519949 &           1.0 &          -1.0 &      -0.025778 &      -0.007938 &       0.044574 &      -0.187635 &      -0.066197 &      -0.519949 \\
      0.264101 &       0.228089 &      -0.095608 &       0.330662 &           1.0 &          -1.0 &      -0.029167 &      -0.108153 &       0.264101 &      -0.228089 &      -0.095608 &      -0.330662 \\
      0.170093 &       0.642822 &      -0.063692 &      -0.628638 &           1.0 &          -1.0 &      -0.613767 &      -0.089652 &       0.170093 &      -0.642822 &      -0.063692 &       0.628638 \\
     -0.149765 &      -0.498075 &       0.098375 &       0.103394 &           1.0 &          -1.0 &       0.167898 &       0.092520 &      -0.149765 &       0.498075 &       0.098375 &      -0.103394 \\
      0.885004 &       0.924225 &      -0.872434 &      -0.489577 &           1.0 &          -1.0 &      -0.824537 &      -0.634052 &       0.885004 &      -0.924225 &      -0.872434 &       0.489577 \\
      0.479148 &       0.807361 &      -0.473696 &       0.863558 &           1.0 &          -1.0 &      -0.201012 &      -0.159771 &       0.479148 &      -0.807361 &      -0.473696 &      -0.863558 \\
      0.531466 &       0.884858 &      -0.814464 &       0.967916 &           1.0 &          -1.0 &       0.819537 &      -0.015970 &       0.531466 &      -0.884858 &      -0.814464 &      -0.967916 \\
      0.613966 &       0.956273 &      -0.923856 &       0.757251 &           1.0 &          -1.0 &       0.026984 &      -0.004039 &       0.613966 &      -0.956273 &      -0.923856 &      -0.757251 \\
      0.589206 &       0.811314 &      -0.285452 &      -0.695849 &           1.0 &           1.0 &      -0.922549 &      -0.252909 &       0.589206 &      -0.811314 &      -0.285452 &       0.695849 \\
      0.894921 &       0.988857 &      -0.820157 &      -0.952284 &           1.0 &          -1.0 &      -0.981286 &      -0.608243 &       0.894921 &      -0.988857 &      -0.820157 &       0.952284 \\
\bottomrule
\end{tabular}

}

\resizebox{\textwidth}{!}{
\begin{tabular}{cccccccccccc}
\toprule
 z\_vx\_eqt\_corr &  z\_vx\_gal\_corr &  z\_vy\_eqt\_corr &  z\_vy\_gal\_corr &  z\_vz\_eqt\_corr &  z\_vz\_gal\_corr &  vx\_vy\_eqt\_corr &  vx\_vy\_gal\_corr &  vx\_vz\_eqt\_corr &  vx\_vz\_gal\_corr &  vy\_vz\_eqt\_corr &  vy\_vz\_gal\_corr \\
\midrule
     -0.025778 &       0.007938 &       0.044574 &       0.187635 &      -0.066197 &       0.519949 &        0.994716 &       -0.979444 &        0.996427 &        0.786199 &        0.993584 &       -0.674004 \\
     -0.029167 &       0.108153 &       0.264101 &       0.228089 &      -0.095608 &       0.330662 &        0.954551 &       -0.942633 &        0.995686 &        0.960283 &        0.934157 &       -0.830872 \\
     -0.613767 &       0.089652 &       0.170093 &       0.642822 &      -0.063692 &      -0.628638 &        0.501601 &       -0.504215 &        0.666283 &        0.433038 &        0.921174 &       -0.798679 \\
      0.167898 &      -0.092520 &      -0.149765 &      -0.498075 &       0.098375 &       0.103394 &        0.941081 &       -0.794564 &        0.969693 &        0.968901 &        0.952575 &       -0.889424 \\
     -0.824537 &       0.634052 &       0.885004 &       0.924225 &      -0.872434 &      -0.489577 &       -0.842508 &        0.664945 &        0.742314 &       -0.497302 &       -0.817323 &       -0.592257 \\
     -0.201012 &       0.159771 &       0.479148 &       0.807361 &      -0.473696 &       0.863558 &        0.745502 &       -0.447692 &        0.939536 &        0.563474 &        0.543182 &        0.438997 \\
      0.819537 &       0.015970 &       0.531466 &       0.884858 &      -0.814464 &       0.967916 &        0.844144 &       -0.406029 &       -0.402566 &        0.160099 &        0.040033 &        0.785306 \\
      0.026984 &       0.004039 &       0.613966 &       0.956273 &      -0.923856 &       0.757251 &        0.791982 &       -0.279000 &        0.344724 &        0.640388 &       -0.271265 &        0.540706 \\
     -0.922549 &      -0.252909 &       0.589206 &      -0.811314 &      -0.285452 &       0.695849 &       -0.299324 &       -0.343299 &        0.547329 &       -0.697581 &        0.590029 &       -0.254861 \\
     -0.981286 &       0.608243 &       0.894921 &       0.988857 &      -0.820157 &      -0.952284 &       -0.827424 &        0.514619 &        0.863911 &       -0.535484 &       -0.505155 &       -0.953220 \\
\bottomrule
\end{tabular}
}

\end{table*}

\section{Transformation from Equatorial to Galactic coordinate system}
\label{sec:e2g}

The following relationship exists between coordinates in Equatorial and Galactic coordinate systems:
\begin{equation}
\begin{aligned}
    &\cos{b}\cos{(l_{\rm NEP}-l)} = \cos{(\delta_{\rm NGP})}\sin{\delta}-\sin{(\delta_{\rm NGP})}\cos{\delta}\cos{(\alpha-\alpha_{\rm NGP})}~,\\
    &\cos{b}\sin{(l_{\rm NEP}-l)} = \cos{\delta}\sin{(\alpha-\alpha_{\rm NGP})}~,\\
    &\sin{b} = \sin{(\delta_{\rm NGP})}\sin{\delta}+\cos{(\delta_{\rm NGP})}\cos{\delta}\cos{(\alpha-\alpha_{\rm NGP})}~.\\
    	\label{eq:gal}
\end{aligned}
\end{equation}

During the calculation process, the galactic pole coordinates are consistent with those in the Python package \emph{astropy}, where

\begin{equation}
\begin{aligned}
    &l_{\rm NGP} = 122.93192526\degree~,\\
    &\alpha_{\rm NGP} = 192.85947789\degree~,\\
    &\delta_{\rm NGP} = 27.12825241\degree ~.
    \label{eq:value}
\end{aligned}
\end{equation}

The position coordinate matrices of coordinates in Equatorial ($\rm eqt$) and Galactic ($\rm gal$) coordinate systems are defined as follows:
\begin{equation}
    \bm{P}^{\rm eqt}=
\begin{bmatrix}
    x^{\rm eqt}\\
    y^{\rm eqt}\\
    z^{\rm eqt}\\
\end{bmatrix}
    =
\begin{bmatrix}
    \cos{\delta}\cos{\alpha}\\
    \cos{\delta}\sin{\alpha}\\
    \sin{\delta}\\
\end{bmatrix}
    ,  
    \bm{P}^{\rm gal}=
\begin{bmatrix}
    x^{\rm gal}\\
    y^{\rm gal}\\
    z^{\rm gal}\\
\end{bmatrix}
    =
\begin{bmatrix}
    \cos{b}\cos{l}\\
    \cos{b}\sin{l}\\
    \sin{b}\\
\end{bmatrix}~.
\end{equation}

By separating the position coordinate matrices from the equations \ref{eq:gal}, we extract the rotation matrix $\bm{R}$:

\begin{equation}
    \bm{R} =
\begin{bmatrix}
    -\sin{(\delta_{\rm NGP})}\cos{(\alpha_{\rm NGP})} & -\sin{(\delta_{\rm NGP})}\sin{(\alpha_{\rm NGP})} & \cos{(\delta_{\rm NGP})}\\
    -\sin{(\alpha_{\rm NGP})} & \cos{(\alpha_{\rm NGP})} & 0\\
    \cos{(\delta_{\rm NGP})}\cos{(\alpha_{\rm NGP})} & \cos{(\delta_{\rm NGP})}\sin{(\alpha_{\rm NGP})} & \sin{(\delta_{\rm NGP})}
\end{bmatrix}~.
\end{equation}

The velocity coordinate matrix $\bm{V}^{\rm eqt}$ is the derivative of the equatorial position coordinate matrix with respect to time. Therefore, for both position and velocity propagation, it just needs to multiply by the rotation matrix as follows:
\begin{equation}
\begin{aligned}
    &\bm{P}^{\rm gal} = \bm{R}\bm{P}^{\rm eqt}~,\\
    &\bm{V}^{\rm gal} = \bm{R}\dot{\bm{P}}^{\rm eqt} = \bm{R}\bm{V}^{\rm eqt}~.
    	\label{eq:galmx}
\end{aligned}
\end{equation}

Hence, the linear and non-linear calculation of galactic coordinates only depends on the matrices of Cartesian coordinates $\bm{P}$ and velocities $\bm{V}$ in the Equatorial coordinate system. The galactic variance-covariance matrix is calculated as follows:
\begin{equation}
    \bm{C}^{\rm gal}_{\rm car} = \bm{R}\bm{C}^{\rm eqt}_{\rm car}\bm{R}^{T}~.
\end{equation}

\section{Jacobian matrix}
\label{j}
The elements of the Jacobian matrix are given below, where $\rm {A}=1\,\rm{AU}$:

\begin{flalign}
&\ \bm{J}_{11} = \frac{\partial\rm x}{\partial\alpha} =  -\frac{\rm{A}\sin{\alpha}\cos{\delta}}{\varpi} &
\end{flalign}

\begin{flalign}
&\ \bm{J}_{12} = \frac{\partial\rm x}{\partial\delta} =  -\frac{\rm{A}\cos{\alpha}\sin{\delta}}{\varpi} &
\end{flalign}

\begin{flalign}
    &\ \bm{J}_{13} = \frac{\partial\rm x}{\partial\varpi} =  -\frac{\rm{A}\cos{\alpha}\cos{\delta}}{\varpi^2} &
\end{flalign}

\begin{flalign}
    &\bm{J}_{14} = \frac{\partial\rm x}{\partial\mu_{\alpha^*}} =  0 &
\end{flalign}
\begin{flalign}
    &\bm{J}_{15} = \frac{\partial\rm x}{\partial\mu_{\delta}} =  0 &
\end{flalign}
\begin{flalign}
    &\bm{J}_{16} = \frac{\partial\rm x}{\partial{v_r}} =  0 &
\end{flalign}

\begin{flalign}
    &\bm{J}_{21} = \frac{\partial\rm y}{\partial\alpha} =  \frac{\rm{A}\cos{\alpha}\cos{\delta}}{\varpi} &
\end{flalign}

\begin{flalign}
    &\bm{J}_{22} = \frac{\partial\rm y}{\partial\delta} =  -\frac{\rm{A}\sin{\alpha}\sin{\delta}}{\varpi} &
\end{flalign}

\begin{flalign}
    &\bm{J}_{23} = \frac{\partial\rm y}{\partial\varpi} =  -\frac{\rm{A}\sin{\alpha}\cos{\delta}}{\varpi^2} &
\end{flalign}
\begin{flalign}
    &\bm{J}_{24} = \frac{\partial\rm y}{\partial\mu_{\alpha^*}} =  0 &
\end{flalign}
\begin{flalign}
    &\bm{J}_{25} = \frac{\partial\rm y}{\partial\mu_{\delta}} =  0 &
\end{flalign}
\begin{flalign}
    &\bm{J}_{26} = \frac{\partial\rm y}{\partial{v_r}} =  0 &
\end{flalign}
\begin{flalign}
    &\bm{J}_{31} = \frac{\partial\rm z}{\partial\alpha} =  0 &
\end{flalign}

\begin{flalign}
    &\bm{J}_{32} = \frac{\partial\rm z}{\partial\delta} =  \frac{\rm{A}\cos{\delta}}{\varpi} &
\end{flalign}

\begin{flalign}
    &\bm{J}_{33} = \frac{\partial\rm z}{\partial\varpi} =  -\frac{\rm{A}\sin{\delta}}{\varpi^2} &
\end{flalign}

\begin{flalign}
    &\bm{J}_{34} = \frac{\partial\rm z}{\partial\mu_{\alpha^*}} =  0 &
\end{flalign}
\begin{flalign}
    &\bm{J}_{35} = \frac{\partial\rm z}{\partial\mu_{\delta}} =  0 &
\end{flalign}
\begin{flalign}
    &\bm{J}_{36} = \frac{\partial\rm z}{\partial{v_r}} =  0 &
\end{flalign}

\begin{flalign}
    &\bm{J}_{41} = \frac{\partial\rm v_x}{\partial\alpha} =  -v_r\sin{\alpha}\cos{\delta}+\frac{\rm{A}\mu_{\delta}\sin{\alpha}\sin{\delta}}{\varpi}-\frac{\rm{A}\mu_{\alpha^*}\cos{\alpha}}{\varpi}&
\end{flalign}

\begin{flalign}
    &\bm{J}_{42} = \frac{\partial\rm v_x}{\partial\delta} = -v_r\cos{\alpha}\sin{\delta}-\frac{\rm{A}\mu_{\delta}\cos{\alpha}\cos{\delta}}{\varpi} &
\end{flalign}

\begin{flalign}
    &\bm{J}_{43} = \frac{\partial\rm v_x}{\partial\varpi} =  \frac{\rm{A}\mu_{\delta}\cos{\alpha}\sin{\delta}}{\varpi^2}+\frac{\rm{A}\mu_{\alpha^*}\sin{\alpha}}{\varpi^2} &
\end{flalign}

\begin{flalign}
    &\bm{J}_{44} = \frac{\partial\rm v_x}{\partial\mu_{\alpha^*}} =  -\frac{\rm{A}\sin{\alpha}}{\varpi} &
\end{flalign}

\begin{flalign}
    &\bm{J}_{45} = \frac{\partial\rm v_x}{\partial\mu_{\delta}} =  -\frac{\rm{A}\cos{\alpha}\sin{\delta}}{\varpi} &
\end{flalign}

\begin{flalign}
    &\bm{J}_{46} = \frac{\partial\rm v_x}{\partial v_r} =  \cos{\alpha}\cos{\delta} &
\end{flalign}

\begin{flalign}
    &\bm{J}_{51} = \frac{\partial\rm v_y}{\partial\alpha} =  v_r\cos{\alpha}\cos{\delta}-\frac{\rm{A}\mu_{\delta}\cos{\alpha}\sin{\delta}}{\varpi}-\frac{\rm{A}\mu_{\alpha^*}\sin{\alpha}}{\varpi} &
\end{flalign}

\begin{flalign}
     &\bm{J}_{52} = \frac{\partial\rm v_y}{\partial\delta} =  -v_r\sin{\alpha}\sin{\delta}- \frac{\rm{A}\mu_{\delta}\sin{\alpha}\cos{\delta}}{\varpi}&
\end{flalign}

\begin{flalign}
    &\bm{J}_{53} = \frac{\partial\rm v_y}{\partial\varpi} =  \frac{\rm{A}\mu_{\delta}\sin{\alpha}\sin{\delta}}{\varpi^2}-\frac{\rm{A}\mu_{\alpha^*}\cos{\alpha}}{\varpi^2} &
\end{flalign}

\begin{flalign}
    &\bm{J}_{54} = \frac{\partial\rm v_y}{\partial\mu_{\alpha^*}} =  \frac{\rm{A}\cos{\alpha}}{\varpi} &
\end{flalign}

\begin{flalign}
    &\bm{J}_{55} = \frac{\partial\rm v_y}{\partial\mu_{\delta}} =  -\frac{\rm{A}\sin{\alpha}\sin{\delta}}{\varpi} &
\end{flalign}

\begin{flalign}
    &\bm{J}_{56} = \frac{\partial\rm v_y}{\partial v_r} =  \sin{\alpha}\cos{\delta} &
\end{flalign}

\begin{flalign}
    &\bm{J}_{61} = \frac{\partial\rm v_z}{\partial\alpha} =  0 &
\end{flalign}

\begin{flalign}
     &\bm{J}_{62} = \frac{\partial\rm v_z}{\partial\delta} =  v_r\cos{\delta}-\frac{\rm{A}\mu_{\delta}\sin{\delta}}{\varpi} &
\end{flalign}

\begin{flalign}
     &\bm{J}_{63} = \frac{\partial\rm v_z}{\partial\varpi} = -\frac{\rm{A}\mu_{\delta}\cos{\delta}}{\varpi^2} &
\end{flalign}

\begin{flalign}
    &\bm{J}_{64} = \frac{\partial\rm v_z}{\partial\mu_{\alpha^*}} =  0 &
\end{flalign}

\begin{flalign}
    &\bm{J}_{65} = \frac{\partial\rm v_z}{\partial\mu_{\delta}} =  \frac{\rm{A}\cos{\delta}}{\varpi} &
\end{flalign}

\begin{flalign}
    &\bm{J}_{66} = \frac{\partial\rm v_z}{\partial v_r} =  \sin{\delta} &
\end{flalign}

%\iffalse
\section{Correlation coefficient}
\label{kij}
In the Eq.~\ref{eq:ccar}, the correlation coefficient of $x$ and $y$ is:

\begin{equation}
    \bm{k}_{\rm xy} =  \frac{\sigma_{xy}}{\sigma_x \sigma_y} = \frac{\sum_{j=1}^{6}\sum_{i=1}^{6}\frac{\partial{\rm{F_x}}}{\partial{b_i}}\frac{\partial{\rm{F_y}}}{\partial{b_j}}\sigma_{b_i}\sigma_{b_j}k_{ij}}{\sqrt{\sum_{i=1}^{6} (\frac{\partial{\rm{F_x}}}{\partial{b_i}}\sigma_{b_i})^2}\sqrt{\sum_{j=1}^{6} (\frac{\partial{\rm{F_y}}}{\partial{b_j}}\sigma_{b_j})^2}}~.
\end{equation}
This equation for $x$ and $y$ is only related to $\alpha$, $\delta$ and $r$ (see Eq.~\ref{eq:carx}, and $r = \frac{\rm A}{\varpi}~$, where $\rm {A}=1\,\rm{AU}$), so $b_i,b_j = \alpha, \delta, \varpi$. Because $\sigma_{\alpha}, \sigma_{\delta}$ is too small compared to the values of $\alpha$ and $\delta$, as $\sigma_{\alpha}, \sigma_{\delta}$ approaches 0, the correlation coefficient of $x$ and $y$ simplifies to:
\begin{equation}
    \bm{k}_{\rm xy} \approx  \frac{\frac{\partial{\rm{F_x}}}{\partial{\varpi}}\frac{\partial{\rm{F_y}}}{\partial{\varpi}}\sigma_{\varpi}^2}{(\frac{\partial{\rm{F_x}}}{\partial{\varpi}}\sigma_{\varpi})(\frac{\partial{\rm{F_y}}}{\partial{\varpi}}\sigma_{\varpi})}~.
\end{equation}

The same is true for the correlation coefficients $k_{xz}$ and $k_{yz}$. As a result, there are many $x$, $y$ and $z$ correlation coefficients approaching $\pm~1$ in the resulting Cartesian catalog. 

%\iffalse
\begin{figure}
\centering
\includegraphics[width=0.75\linewidth]{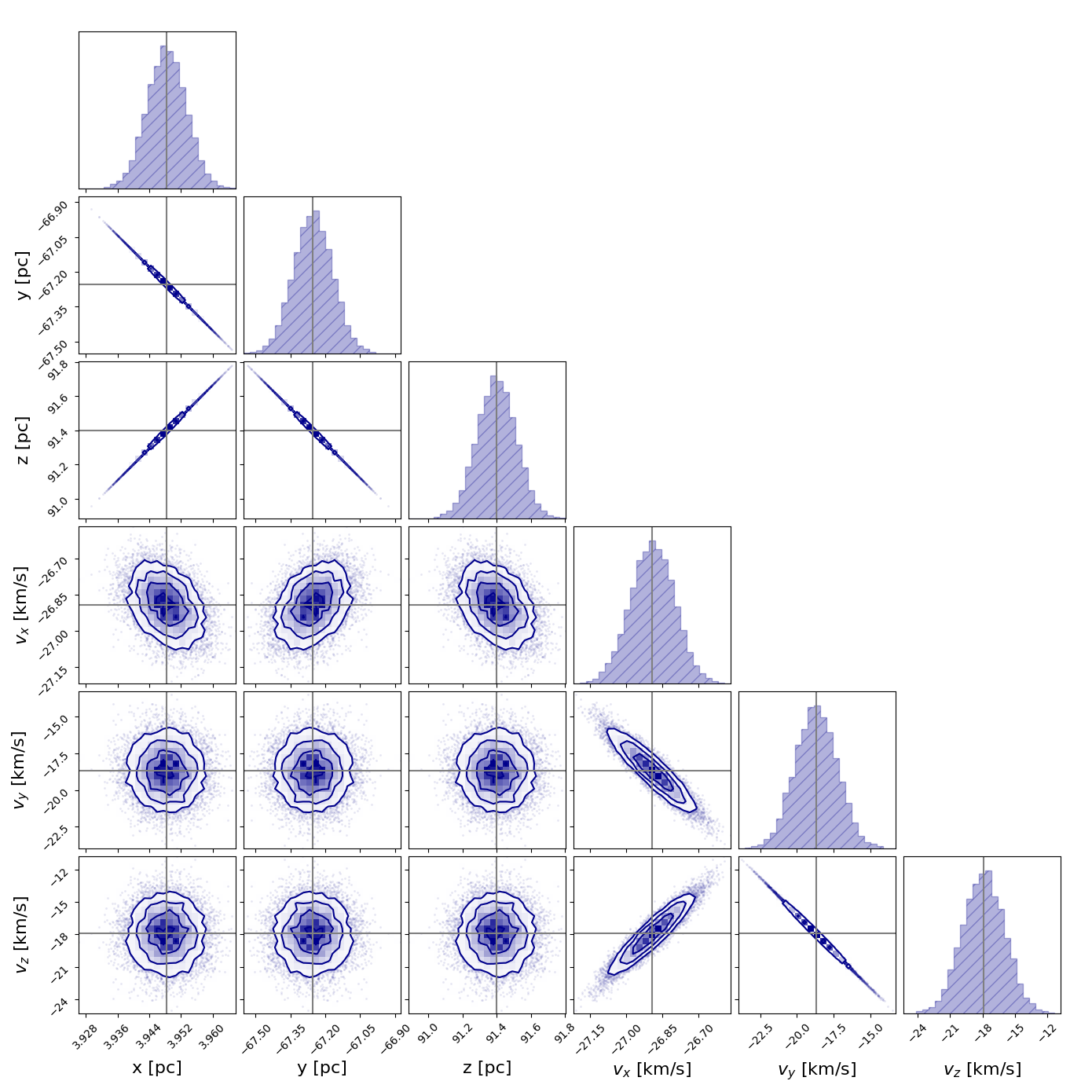}
\includegraphics[width=0.75\linewidth]{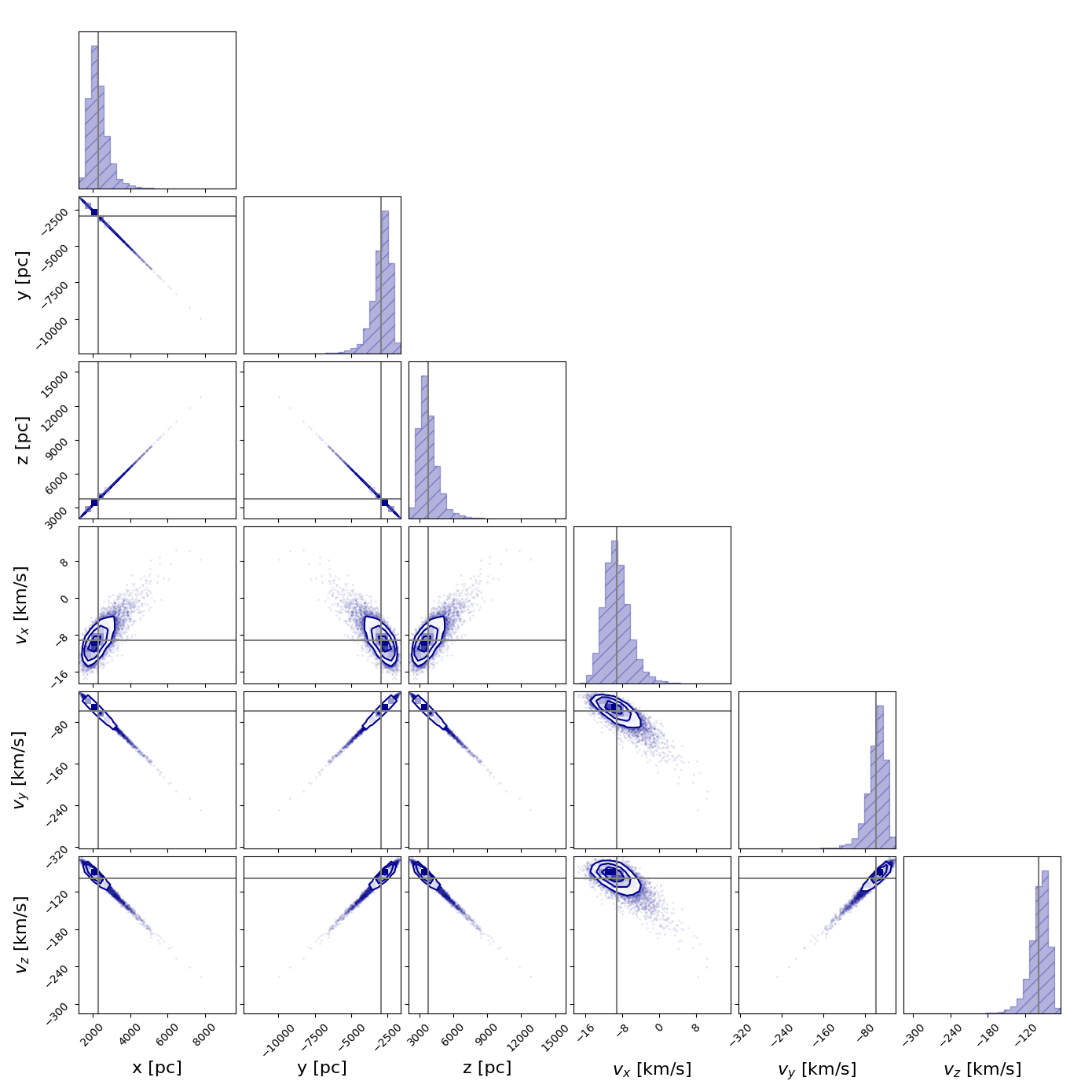}
\caption{Cornerplots showing the distributions of the MC4 method. The MC4 distribution for the equatorial Cartesian coordinates for \emph{Gaia DR3 2149392370122774528} (top) and \emph{Gaia DR3 2070799897454308224} (bottom), which have fractional parallax errors of 0.001 and 0.199, respectively. Blue dots represent MC4 samples while the contours represent 0.5, 1, 1.5 and 2 $\sigma$ confidence levels. The mean is shown by horizontal and vertical grey lines. }
\label{fig:mc}
\end{figure}

Fig.~\ref{fig:mc} shows the sample distributions from the MC4 method. For \emph{Gaia DR3 2149392370122774528} and \emph{Gaia DR3 2070799897454308224}, the correlation coefficient between \(x\), \(y\), and \(z\) is approximately $\pm~1$, regardless of fractional parallax errors. Higher fractional parallax errors lead to a more skewed distribution of the six-dimensional parameters, indicating the significant impact of fractional parallax error on higher-order moments. 
 %\fi

\section{Acronyms and variables}
\label{Acronyms}
For clarity, the de nitions of all key variables and
abbreviations referenced in this work are provided in Appendix \ref{Acronyms}, Table \ref{tab:glossary}.
\begin{table}
 \caption{Glossary of main acronyms and variables.}
 \label{tab:glossary}
 \begin{tabular}{lll}
  \hline
  Symbol & Definition\\
  \hline
  $\alpha$ & Right ascension\\
  $\delta$ & Declination\\
  $\varpi$ & Parallax\\
  $\mu_{\alpha^*}$ & Proper motion in Right ascension, $\mu_{\alpha^*} = \mu_{\alpha} \cos{\delta}$ \\
  $\mu_{\delta}$ & Proper motion in Declination\\
  $v_{r}$ & radial velocity\\
  $k_{ij}$ & Correlation coefficient between i and j\\
  $\bm{C}$ & Variance-Covariance matrix\\
  $\bm{H}$ & Hessian matrix\\
  tr($\bm{I}$) & Trace of matrix $\bm{I}$\\
  $\rm MCi$ & MC error propagation method in sample of $10^i$\\
  $\rm {1st}$ & Linear (first-order) propagation\\
  $\rm {2nd}$ & Second-order error propagation\\
  $r$  & distance\\
  $\sigma_{i}$ & Error (uncertainty) of i\\
  $M_G$ & Absolute magnitude in the G-band\\
  $G_{BP}-G_{RP}$ & BP-RP colour\\
  $l_{\rm NGP}$ & Galactic longitude of the north galactic pole\\
  $l_{\rm NEP}$ & Galactic longitude of the north equatorial pole\\
  $b$ & Galactic latitude\\
  $\rm car$ & Cartesian coordinates\\
  $\rm sph$ & Spherical coordinates\\
  $\rm eqt$ & Equatorial coordinate system\\
  $\rm gal$ & Galactic coordinate system\\

  \hline
 \end{tabular}
\end{table}

\label{lastpage}

\bibliography{bibtex}
\bibliographystyle{raa}
\end{document}